\documentclass{bioinfo}
\copyrightyear{2011}
\pubyear{2011}

\usepackage{url}
\usepackage{subfigure,graphicx}
\usepackage{multirow}
\usepackage{rotating}
\usepackage{epstopdf}
\usepackage{footmisc}

\begin{document}
\firstpage{1}

\title[Evaluating sources of variability in pathway profiling]{EVALUATING SOURCES OF VARIABILITY IN PATHWAY PROFILING}
\author[Barla \textit{et~al}]{
Annalisa Barla$^{(1) \footnote{These authors equally contributed to the work}}$, 
Samantha Riccadonna$^{(2)^*}$, 
Salvatore Masecchia$^{(1)}$, 
Margherita Squillario$^{(1)}$, 
Michele Filosi$^{(2,3)}$, 
Giuseppe Jurman$^{(2)}$ and
Cesare Furlanello$^{(2)}$\footnote{to whom correspondence should be addressed}}
\address{$^{1}$DISI, University of Genoa, via Dodecaneso 35, I-16146 Genova, Italy.\\
$^{2}$FBK, via Sommarive 18, I-38123 Povo (Trento), Italy.\\
$^{3}$CIBIO, University of Trento, via Delle Regole 101, I-38123 Mattarello (Trento), Italy
}

\history{Received on XXXXX; revised on XXXXX; accepted on XXXXX}

\editor{Associate Editor: XXXXXXX}

\maketitle

\begin{abstract}
\section{Motivation:}
A bioinformatics platform is introduced aimed at identifying models of disease-specific pathways, as well as a set of network measures that can quantify changes in terms of global structure or single link disruptions.The approach integrates a network comparison framework with machine learning molecular profiling. ÊThe platform includes different tools combined in one Open Source pipeline, supporting reproducibility of the analysis.
We describe here the computational pipeline and explore the main sources of variability that can affect the results, namely the classifier, the feature ranking/selection algorithm, the enrichment procedure, the inference method and the networks comparison function.

\section{Results:}
The proposed pipeline is tested on a microarray dataset of late stage Parkinsons' Disease patients together with healty controls.
Choosing different machine learning approaches we get low pathway profiling overlapping in terms of common enriched elements.
Nevertheless, they identify different but equally meaningful biological aspects of the same process, suggesting the integration of information across different methods as the best overall strategy.

\section{Availability:}
All the elements of the proposed pipeline are available as Open Source Software: availability details are provided in the main text.

\section{Contact:} \href{annalisa.barla@unige.it}{annalisa.barla@unige.it}
\end{abstract}
\section{Introduction}

We present a computational framework for the study of reproducibility
in network medicine studies \citep{barabasi11network}. Networks,
molecular pathways in particular, are increasingly looked at as a
better organized and more rich version of gene signatures. However,
high variability can be injected by the different methods that are
typically used in system biology to define a cellular wiring diagram
at diverse levels of organization, from transcriptomics to signalling,
of the functional design. For example, to identify the link between
changes in graph structures and disease, we choose and combine in a
workflow a classifier, the feature ranking/selection algorithm, the
enrichment procedure, the inference method and the networks comparison
function. Each of these components is a potential source of
variability, as shown in the case of biomarkers from microarrays
\citep{maqc10maqcII}.  Considerable efforts have been directed to
tackle the problem of poor reproducibility of biomarker signatures
derived from high-throughput -omics data \citep{maqc10maqcII},
addressing the issues of selection bias \citep{ambroise02selection,
  furlanello03entropy} and more recently of pervasive batch effects
\citep{leek10tackling}. We argue that it is now urgent to adopt a
similar approach for network medicine studies. Stability (and thus
reproducibility) in this class of studies is still an open problem
\citep{baralla09inferring}. Underdeterminacy is a major issue
\citep{desmet10advantages}, as the ratio between network dimension
(number of nodes) and the number of available measurements to infer
interactions plays a key role for the stability of the reconstructed
structure. Furthermore, the most interesting applications are based on
inferring networks topology and wiring from high-throughput noisy
measurements \citep{he09reverse}.
\par
Despite its common use even in biological contexts
\citep{sharan06modeling}, the problem of quantitatively comparing
networks (\textit{e.g.}, using a metric instead of evaluating network
properties) is a still an open issue in many scientific disciplines. 
The central problem is of course which network metrics
should be used to evaluate stability, whether focusing on local
changes or global structural changes.  As discussed in
\citet{jurman10introduction}, the classic distances in the edit family
focus only on the portions of the network interested by the
differences in the presence/absence of matching links. Spectral
distances - based on the list of eigenvalues of the Laplacian matrix
of the underlying graph - are instead particularly effective for
studying global structures. In particular, the Ipsen-Mikhailov
\citep{ipsen02evolutionary} distance was found robust in a wide range
of situations \citep{jurman10introduction}. However, global distances
can be tricked by isomorph or close to isomorph graphs. In
\cite{jurman12glocal}, both approaches are improved by proposing a
{\em glocal} measure which combines a spectral distance with a typical
Hamming local editing component. In this paper we use this new tool to
quantify how stability of network reconstruction is modified in
practice by the different inference and enrichment methods.
\par
Pathway enrichment methods are widely used in bioinformatics analysis,
for example to assess the relevance of biomarker lists or as a first
step in network analysis. The enrichment step is performed using the
functional information stored in the Kyoto Encyclopedia of Genes and
Genomes (KEGG) database \citep{kanehisa00KEGG} and in the Gene
Ontology (GO) database \citep{Ashburner:2000}. The reconstruction of
molecular pathways from high-throughput data is then based on the
theory of complex networks (\textit{e.g.}  \citet{strogatz01exploring,
  newman03structure, boccaletti06complex, newman10networks,
  buchanan10networks}) and, in particular, in the reconstruction
algorithms for inferring networks topology and wiring from data
\citep{he09reverse}.
\par
The stability analysis is applied to networks identifying common and
specific network substructures that could be basically associated to
disease. The overall goal of the pipeline is to identify and rank the
pathways reflecting major changes between two conditions, or during a
disease evolution. We start from a profiling phase based on
classifiers and feature selection modules organized in terms of
experimental procedure by Data Analysis Protocol \citep{maqc10maqcII},
obtaining a ranked list of genes with the highest discriminative power.
Classification tasks as well as quantitative phenotype targets can be
considered by using different machine learning methods in this first
phase. Alternative methods are made available as components of one
Open Source pipeline system. The problem of underdeterminacy in the
inference procedure is avoided by focusing only on subnetworks, and
the relevance of the studied pathways for the disease is judged in
terms of discriminative relevance for the underlying classification
problem.
\par
As a testbed for the {\em glocal} stability analysis, we compare network
structures on a collection of microarray data for patients affected by
Parkinson's disease (PD), together with a cohort of controls
\citep{zhang05PD}. PD is a neurodegenerative disorder that impairs the
motor skills at the onset and the cognitive and the speech functions
successively. 
The goal of this task  is to identify the most relevant dirsupted pathways and genes in late stage PD.
On this dataset, we show that choosing different profiling approaches
we get low overlapping in terms of common enriched pathways found.
Despite this variability, if we consider the most disrupted pathways,
their {\em glocal} distances (between case and control networks) share
a common distribution assessing equal meaningfulness to pathways
found starting from different approaches.

\begin{methods}
\section{Materials and Methods}
\begin{figure*}[!tpb]%
\centering
\includegraphics[width=0.95\textwidth]{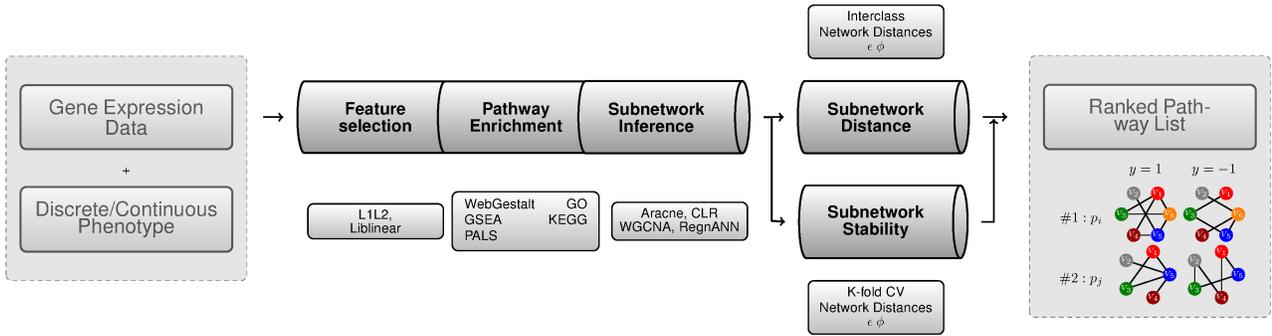}
\caption{General scheme of the analysis pipeline, with the indication of the
algorithms and tools used in the PD dataset application (lower boxes).}
\label{fig:pipeline}
\end{figure*}

The machine learning pipeline adopted in this paper has been originally
introduced in \citep{barla11machine}.
As shown in Figure~\ref{fig:pipeline}, it handles case/control transcription
data through four main steps, 
from a profiling task to the identification of discriminant pathways.
The pipeline is independent from the algorithms used: here we describe each
step and the implementation adopted for the following experiments evaluating
the impact of different enrichment methods in pathway profiling.
       
Formally, we are given a collection of $n$ subjects, each described by a
$p$-dimensional vector $x$ of measurements.
Each sample is also associated with a phenotypical label, e.g. $y \! =\! \{1,-1\}$,
assigning it to a class (in a classification task).
The dataset is therefore represented by a $n\! \times \! p$ expression data
matrix $X$, where $p\! \gg \! n$,  and a corresponding labels vector $Y$.

\noindent\textbf{Feature Selection Step}\noindent\\
The matrix $X$ is used to feed the profiling part of the pipeline
within a proper Data Analysis Protocol, which will ensure accurate and
reproducible results \citep{maqc10maqcII}. 
The prediction model $\mathcal{M}$ is built by using two different algorithms for classification and
feature ranking.
The more recent one is the $\ell_1\ell_2$ regularization with double
optimization, capable of selecting subsets of discriminative genes.  The
algorithm can be tuned to give a minimal set of discriminative genes
or larger sets including correlated genes and it is based on the
optimization principle presented in \cite{zou05regularization}.  The
 implementation used consists of two stages \citep{DeMol09b} and it is cast in nested loops of $10$-fold cross-validation \citep{Barla:2008ESANN}.
The first stage identifies the minimal set of relevant variables (in terms of
prediction error), while, starting from the minimal list, the second one selects
the family of (almost completely) nested lists of relevant variables for
increasing values of linear correlation.
As alternative choice we consider Liblinear, a linear Support Vector Machine (SVM) classifier specifically designed for large datasets (millions of instances and features) \citep{fan08liblinear}.
In particular, the classical dual optimization problem with L2-SVM loss function is solved with a coordinate descent method. For our experiment we adopt the $\ell_2$-regularized penalty term and the module of the weights for ranking purposes within a $100\times 3$-fold cross validation schema. 
We build a model for increasing feature sublists where the feature ranking is defined according to the importance for the classifier. We choose the model, and thus the top ranked features, providing a balance between the accuracy of the
classifier and the stability of the signature \citep{maqc10maqcII}.
Thus, the output of this first  step is a gene signature $g_1, ..., g_k$ 
(one for each model $\mathcal{M}$) containing the $k$ most discriminant features, ranked according their frequency score.

\noindent\textbf{Pathway Enrichment}\noindent\\
The successive enrichment phase derives a list of relevant pathways from the
discriminant features, moving the focus of the analysis from single genes to
functionally related pathways. 
As outlined in the review by \citet{huang09bioinformatics}, in the last 10
years the gene-annotation enrichment analysis field has been growing rapidly
and several bioinformatics tools have been designed for this task.
\citet{huang09bioinformatics} provide a unique categorization of these
enrichment tools in three major categories based on the underlying algorithm:
singular enrichment analysis (SEA), gene set enrichment analysis (GSEA),
and modular enrichment analysis (MEA).
We choose one representative $\mathcal{E}$ for each class for our comparison referring as sources of information 
$\mathcal{D}$ both to the KEGG, to explore known information on 
molecular interaction networks, 
and GO, to explore functional characterization and biological annotation. 
In the first category we choose WebGestalt (WG),
an online gene set analysis toolkit \citep{zhang05webgestalt} taking as input
a list of relevant genes/probesets. The enrichment analysis is performed in
KEGG and GO identifying the most relevant pathways and ontologies in the
signatures.  WG adopts the hypergeometric test to evaluate functional category enrichment and performs a multiple test adjustment (the default method is the one from \cite{Benjamini:1995}).
The user may choose different significance levels and the minimum number of genes belonging to the selected functional groups.
GSEA \citep{subramanian05gene} is our representative of the second class. 
It first performs a correlation analysis between the features and the
phenotype obtaining a ranked list of features.
Secondly it determines whether 
the members of given gene sets are randomly distributed in the ranked list of features obtained above, or primarily found at the top or bottom.
We use the {\em preranked} analysis tool, feeding the  ranked lists of genes produced by the profiling phase directly to the enrichment step of GSEA. 
To avoid a miscalculation of the enrichment score ES, we provide as input the complete list of variables (not just the selected ones), assigning to the not-selected a zero score. 
Note that GSEA calcultates enrichment scores that reflect the degree to which a 
pathway is overrepresented at the top or the bottom of the ranked list. 
In  our analysis we considered only pathways enriched with the top of the list. 
Finally, the tool in the MEA class is the Pathways and Literature Strainer (PaLS) 
\citep{alibes08PALS}, which takes a list or a set of lists of genes (or protein
identifiers) and shows which ones share the same GO terms or KEGG pathways, 
following a criterion based on a threshold {\em t} set by the user. 
The tool provides as output those functional groups that are shared at least by the {\em t$\%$} of the selected genes.
PaLS is aimed at easing the biological interpretation of results from  studies of differential expression and gene selection, without assigning any statistical significance to the final output. 
Applying the above mentioned pathway enrichment techniques, we retrieve for
each gene $g_i$ the corresponding whole pathway $p_i = \{h_1,...,h_t\}$, where
the genes $h_j \neq g_i$ not necessarily belong to the original signature
$g_1, ..., g_k$. Extending the analysis to all the $h_j$ genes of the pathway
allows us to explore functional interactions that would otherwise get lost.

\noindent\textbf{Subnetwork Inference}\noindent\\
For each pathway, networks are inferred separately on data from the different
classes.
The subnetwork inference phase requires to reconstruct a network $N_{p_i,y}$ 
on the pathway $p_i$ by using the steady state expression data of the samples of each
class $y$. The network inference procedure is limited to the sole genes
belonging to the pathway $p_i$ in order to avoid the problem of intrinsic
underdeterminacy of the task. As an additional caution against this problem,
in the following experiments we limit the analysis to pathways having more than
$4$ nodes and less than $1000$ nodes. 
The pipeline allows to run analysis in parallel with different methods and thus to evaluate the variability along the whole pipeline.
We adopt four different subnetwork
reconstruction algorithms $\mathcal{N}$: the Weighted Gene Co-Expression Networks (WGCN)
algorithm \citep{horvath11weighted}, the Algorithm for the Reconstruction of
Accurate Cellular Networks (ARACNE) \citep{margolin06aracne}, the
Context Likelihood of Relatedness (CLR) approach \citep{faith07large}, and the Reverse Engineering Gene Networks using Artificial Neural Networks (RegnANN) \citep{grimaldi11regnann}.
In this work, we applied WGCNA, CLR and ARACNE to analyze the pathway identified in the Pathway Enrichment step, while 
RegnANN was used, as an alternative algorithm, to reconstruct interesting disrupted pathways 
and to compare its results with results from methods mentioned above.
WGCNA is based on the idea of using (a function of) the absolute
correlation between the expression of a couple of genes across the samples to
define a link between them. 
ARACNE is a method for inferring networks from the transcription level
\citep{margolin06aracne} to the metabolic level \citep{nemenman07reconstruction}.
Beside it was originally designed for handling the complexity of regulatory
networks in mammalian cells, it is able to address a wider range of network
deconvolution problems.
This information-theoretic algorithm removes the vast majority of indirect
candidate interactions inferred by co-expression methods by using the data
processing inequality property \citep{cover91elements}.
CLR belongs to the relevance networks class of algorithms and is employed for the identification of transcriptional regulatory interactions \citep{faith07large}. 
In particular, interactions between transcription factors and gene targets are scored by using the mutual information between the corresponding gene expression levels coupled with an adaptive background correction step. Indeed the most probable regulator-target interactions are chosen comparing the mutual information score versus the "background" distribution of mutual information scores for all possible pairs within the corresponding network context (\textit{i.e.} all the pairs including either the regulator or the target).
RegnANN is a newly defined method for inferring gene regulatory networks based on an ensemble of feed-forward multilayer perceptrons. Correlation is used to define gene interactions. For each gene a one-to-many regressor is trained using the transcription data to learn the relationship between the gene and all the other genes of the network. The interaction among genes  are estimated independently and the overall network is obtained by joining all the neighborhoods.
Summarizing, we obtain a real-valued adjacency matrix as output of the subnetwork inference step for each dataset $X$, for each class $y$, for each model $\mathcal{M}$, for each enrichment tool $\mathcal{E}$, for each source of information $\mathcal{D}$, for each pathway $p_i$, and for each subnetwork inference algorithm $\mathcal{N}$.
We thus need to quantitatively evaluate network differences, \textit{i.e.} using a metric instead of evaluating network properties. 

\noindent\textbf{Subnetwork Distance and Stability}\noindent\\
Among the possible choices already available in literature, we focus on two of the most common distance families: the set of edit-like distances and the spectral distances.
The functions in the former family quantitatively evaluate the differences between two networks (with the same number of nodes) in terms of minimum number of edit operations (with possibly different costs) transforming one network into the other, \textit{i.e.} deletion and insertion of links, while spectral measures relies on functions of the eigenvalues of one of the connectivity matrices of the underlying graph.
As discussed in \cite{jurman10introduction}, the drawback of many classical distances (such as those of the edit family) is locality, that is focusing only on the portions of the network interested by the differences in the presence/absence of matching links.  
Spectral distances can overcome this problem considering the global structure of the compared topologies. 
Within them, we consider the Ipsen-Mikhailov $\epsilon$ distance: originally introduced in \cite{ipsen02evolutionary} as a tool for network reconstruction from its Laplacian spectrum, it has been proven to be the most robust in a wide range of situations by \cite{jurman10introduction}.
We are also aware that spectral measures are not flawless: they cannot distinguish isomorphic or isospectral graphs, which can correspond to quite different conditions within the biologica context.
We thus introduce the \textit{glocal} distance $\phi$ as a possible solution against both issues: $\phi$ is defined as the product metric of the Hamming distance H (as representative of the edit-familiy) and the $\epsilon$ distance.
Full mathematical details are available in \cite{jurman12glocal}.

Relying on the distances $\epsilon$ and $\phi$, we evaluate networks corresponding to the same pathway 
for different classes, \textit{i.e.} all the pairs $(N_{p_i,+1},N_{p_i,-1})$ and rank the pathways 
themselves from the most to the least changing across classes. 

Moreover, we attached to each
network a quantitative measure of stability with respect to data subsampling,
in order to evaluate the reliability of inferred topologies. 
In particular, for each $N_{p_i,y}$, we extracted a random subsampling
(of a fraction $r$ of $X$ labelled as $y$) on which the corresponding $N_{p_i,y}$
will be reconstructed.
Repeating $m$ times the subsampling/inferring procedure, a set of
$m$ nets will be generated for each $N_{p_i,y}$.
Then all mutual $\binom{m}{2}=\frac{m(m-1)}{2}$ distances are computed, and for each set of $m$ graphs
we build the corresponding distance histogram. 
In particular, for our experiments we set $m=20$ and $r=\frac{2}{3}$.
Mean and variance of the constructed histograms will quantitatively assess the
stability of the subnetwork inferred from the whole dataset: the lower the
values, the higher the stability in terms of robustness to data perturbation
(subsampling).
%

\noindent
\textbf{Data description and preprocessing}\noindent \\
The presented approach is applied to PD data originally introduced
in \cite{zhang05PD} and publicly available at Gene Expression Omnibus (GEO),
with accession number GSE20292.  The biological samples consist of whole
substantia nigra tissue in $11$ PD patients and $18$ healthy controls.
Expressions were measured on Affymetrix HG-U133A platform. We perform the data
normalization on the raw data with the {\em rma} algorithm of the R Bioconductor {\em affy} package 
with a custom CDF (downloaded from BrainArray: \href{http://brainarray.mbni.med.umich.edu}{http://brainarray.mbni.med.umich.edu}) adopting Entrez identifiers. 

\noindent
\textbf{Software Availability}\noindent \\
The Python implementation of $\ell_1\ell_2$ regularization with double
optimization is available at  \href{http://slipguru.disi.unige.it/Software/L1L2Py}{http://slipguru.disi.unige.it/Software/L1L2Py}.
Liblinear was originally developed by the Machine Learning Group at the National Taiwan University
and it is now available within the Python \texttt{mlpy} library (\href{http://mlpy.fbk.eu}{http://mlpy.fbk.eu}). 
We adopt the \texttt{l2r\_l2loss\_svc\_dual} solver, with $C\!=\!10e^{-4}$.
WG is available as a web application at  {\href{http://bioinfo.vanderbilt.edu/webgestalt}{http://bioinfo.vanderbilt.edu/webgestalt}}.
GSEA is available either as a web application or a Java stand-alone tool at 
\href{http://www.broadinstitute.org/gsea}{http://www.broadinstitute.org/gsea}.
PaLS is available online at \href{http://pals.bioinfo.cnio.es}{http://pals.bioinfo.cnio.es} as a web application.                
For three of the network reconstruction algorithms, we adopted the R Bioconductor implementation: the \textit{WGCNA} package for WGCN, and
\textit{MiNET (Mutual Information NETworks package)} for ARACNE and CLR.
In particular, we set the WGCNA soft thresholding exponent to 5, while we keep the default value for the ARACNE data processing
inequality tolerance parameter \citep{meyer08minet}. Moreover, the ARACNE
implementation requires all the features to have non-zero variance on each
class and for consistency purposes we applied this in all experiments.
RegnANN is instead available from \href{http://sourceforge.net/projects/regnann}{http://sourceforge.net/projects/regnann}.
It is implemented in C and relies on GPGPU programming paradigm for improving efficiency.
The {\em glocal} distance $\phi$ is available upon request to the authors either as R script or Python script. The computation of the Ipsen-Mikhailov distance $\epsilon$ is included as component of the {\em glocal} script.

\end{methods}

\section{Results and Discussion}
\begin{table}[!t]
\processtable{Summary of pathways retrieved in the pathway enrichment step. The numbers in brackets refer to the pathways  considered for the  network inference step. 
\label{Tab:NPathway}
}
{\begin{tabular}{ccrrr}\toprule
\multirow{2}{*}{$\mathcal{M}$} & \multirow{2}{*}{$\mathcal{D}$} & \multicolumn{3}{c}{$\mathcal{E}$} \\
                               &                                & \multicolumn{1}{c}{WG}
                                                                & \multicolumn{1}{c}{GSEA}
                                                                & \multicolumn{1}{c}{PaLS}   \\\midrule
\multirow{2}{*}{$\ell_1\ell_2$}& GO   & 114 (92) & 7 (7) & 381 (331) \\
                               & KEGG & 43 (43)  & 2 (2) & \hfill71 \hfill(71)    \\
\multirow{2}{*}{Liblinear}     & GO   & 83 (45)  & 0 (0) & 404 (356)    \\
                               & KEGG & 56 (55)  & 1 (1) & \hfill77 \hfill(77)    \\\botrule
\end{tabular}}{}
\end{table}
The feature selection results varied accordingly to the chosen method: 
$\ell_1\ell_2$ identified $458$ discriminant genes associated to an average prediction performance of $80.8\%$, 
while with Liblinear we selected the top-$500$ genes associated to an accuracy of $80\%$ (95\% boostrap Confidence Interval:  (0.78;0.83)) coupled with a stability of 0.70.
The lists have $119$ common genes. 

The number of enriched pathways greatly varied depending on the selection and
enrichment tools. 
With $\ell_1\ell_2$, we found globally for GO and KEGG, 157, 452 and 9 pathways as significantly enriched, for WG, PaLS and GSEA respectively.
Similarly, for Liblinear, the identified pathways  were: 139, 481 and 1.  Table
\ref{Tab:NPathway} reports the detailed results for model $\mathcal{M}$, enrichment $\mathcal{E}$ and database $\mathcal{D}$.
\begin{figure}[tb]%
\begin{center}
\begin{tabular}{c}
\includegraphics[width=0.35\textwidth]{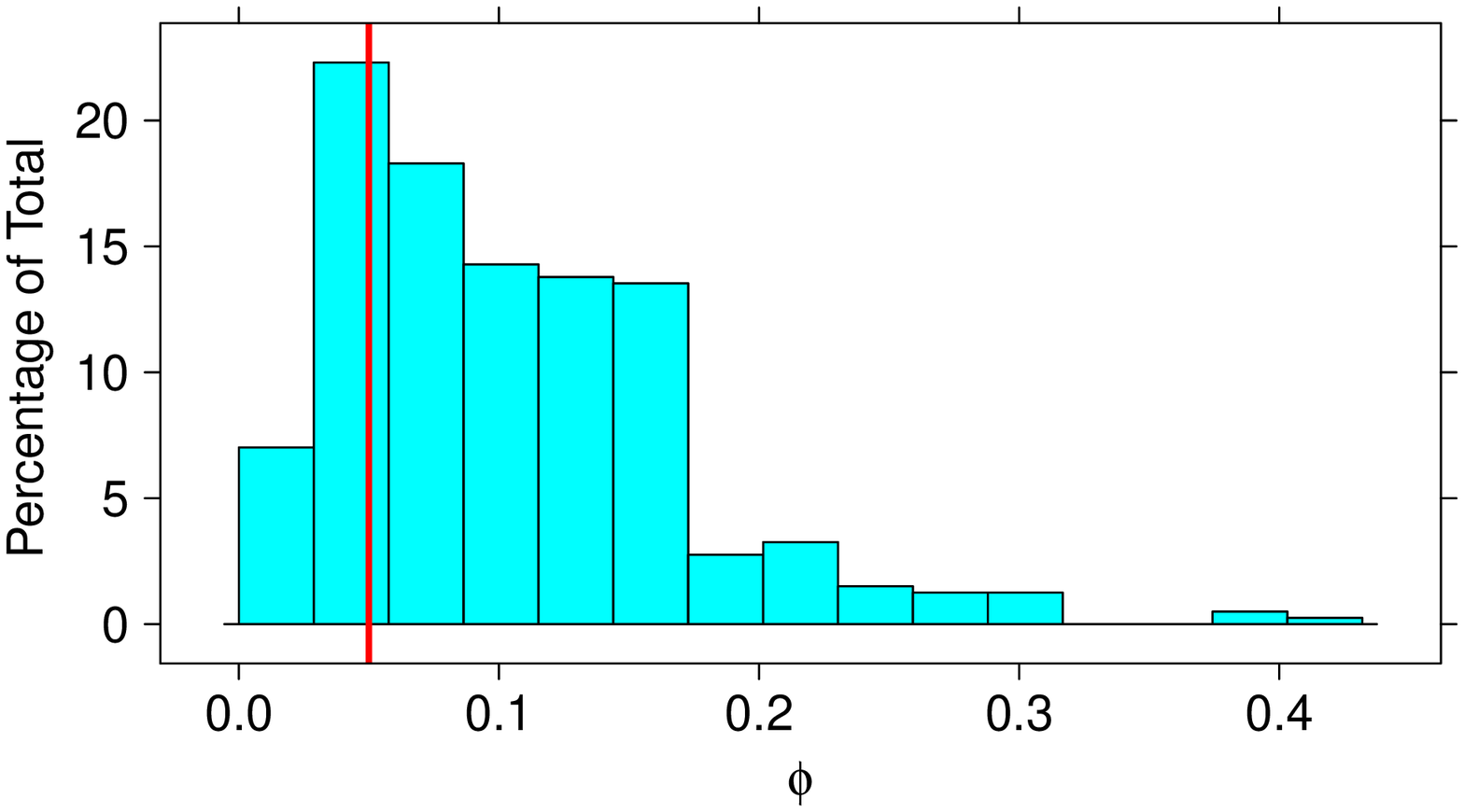} \\
(a)\\
\includegraphics[width=0.35\textwidth]{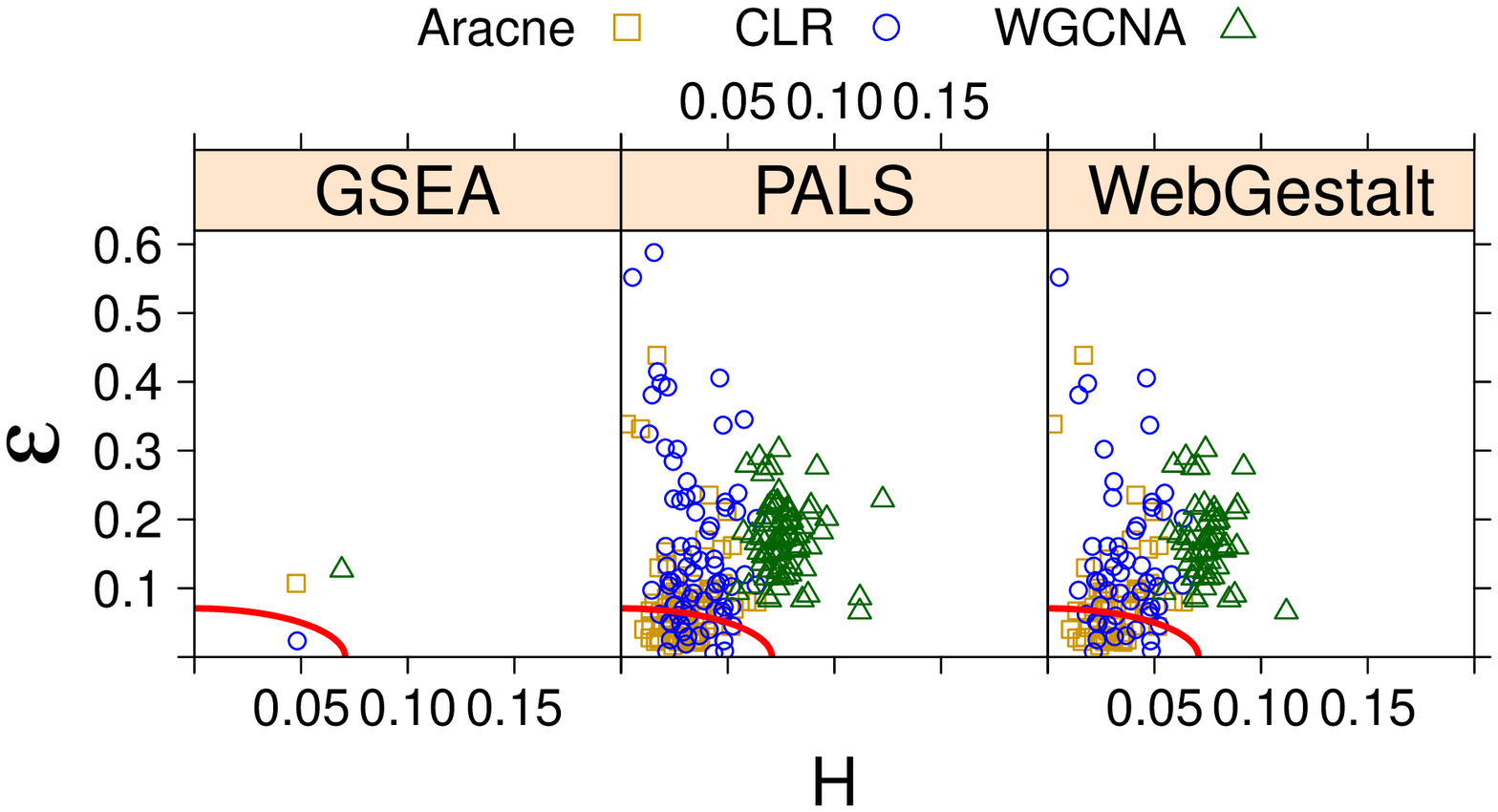} \\
(b)\\
\includegraphics[width=0.4\textwidth]{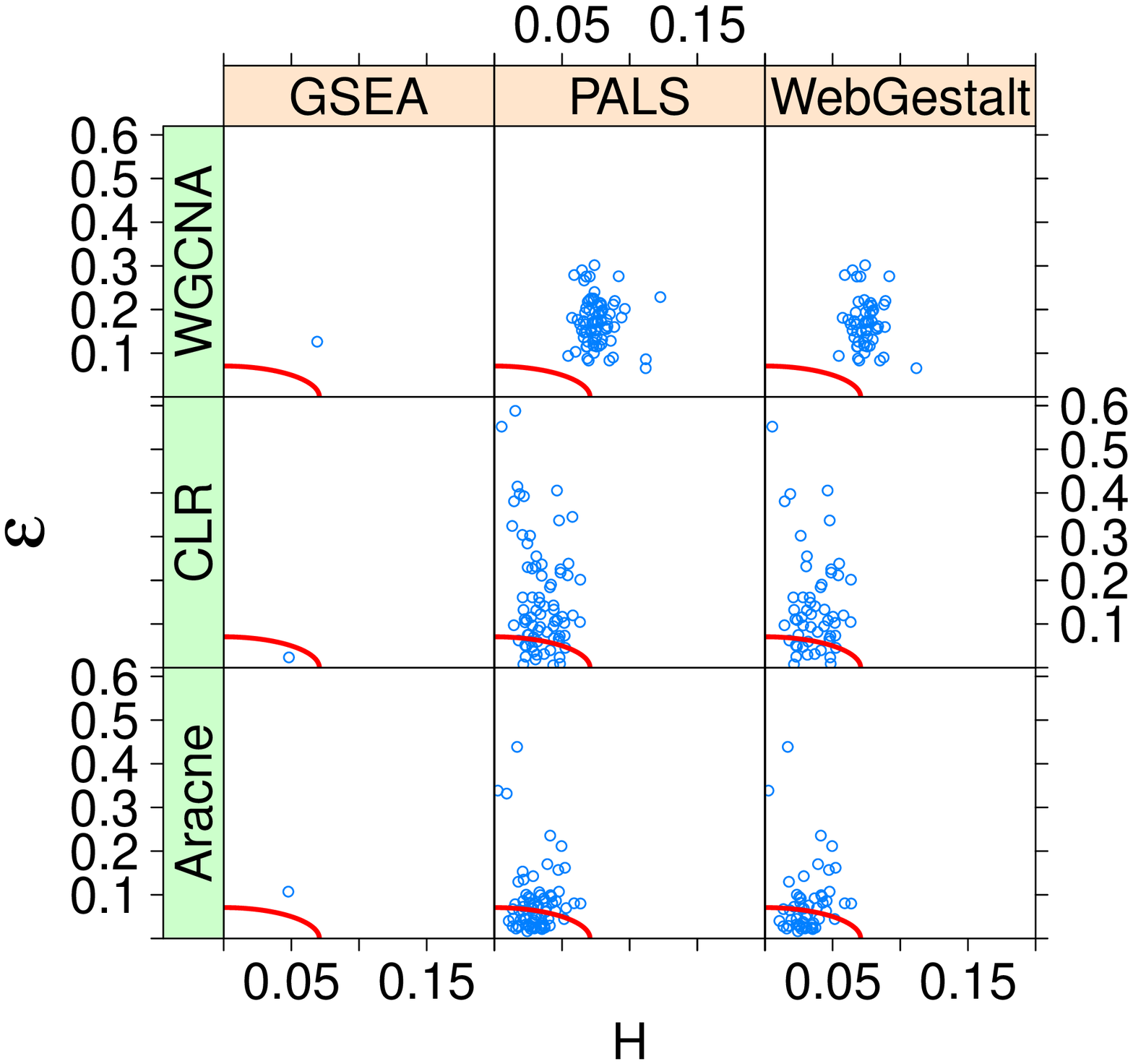} \\
(c)\\
\end{tabular}
\end{center}
\caption{Detailed distance plot for Liblinear and KEGG (see Figure \ref{Fig:HammingvsIpsen}b). 
The histogram plot in (a) represents the cumulative histogram for all distances across enrichment methods $\mathcal{E}$ and
subnetwork inference algorithms $\mathcal{N}$. The threshold $\tau$ is set to retain at least $50\%$ of pathways. In (b) a plot of 
Hamming vs. Ipsen distances (H vs. $\epsilon$) for all possible combinations of $\mathcal{E}$ and $\mathcal{N}$,
which is detailed in (c).
}
\label{Fig:HammingvsIpsenZOOM}
\end{figure}

If we consider the $\ell_1\ell_2$ selection method and the enrichment performed within the GO, 
we may note that no common GO terms were selected across enrichment methods. 
A significant overlap of results was found only between WG and PaLS, with $30$ GO common 
terms.
Similar considerations may be drawn with the results from the Liblinear feature selection method.
Within the GO enrichment we did not identify any common GO term among the three enrichment tools. 
Considering only WG and PaLS, we were able to select $12$ common GO terms.

If we consider the $\ell_1\ell_2$ selection method and the enrichment performed within KEGG, 
two common pathways are identified across enrichment methods.
A significant overlap of results was found between WG and PaLS, with $43$ common pathways.
For Liblinear, only one common pathway was selected among the three enrichment tools.
A significant overlap of results was found between WG and PaLS, with $55$ common pathways.

Following the pipeline, we also performed a comparison of  the three network reconstruction methods.
We considered the most disrupted networks, keeping for the analysis those pathways that had a {\em glocal} distance greater or equal to the chosen threshold $\tau=0.05$.
The choice of such threshold was made considering the distribution of {\em glocal} distances $\phi$ for the methods $\mathcal{M}$. 
For instance, if we consider the Liblinear selection method and the KEGG database, we have a cumulative distribution as depicted in Figure \ref{Fig:HammingvsIpsenZOOM}(a).
The threshold $\tau$ is set to $0.05$ and allows retaining at least $50\%$ of pathways. 
The plot in  Figure \ref{Fig:HammingvsIpsenZOOM}(b) represents the {\em glocal} distances distribution for all enrichment methods $\mathcal{E}$ 
with respect to the two components of the {\em glocal} distance: the Ipsen distance $\epsilon$ and the Hamming distance H. 
The red curved line represents the threshold $\tau$ in this space.  
The plot in Figure \ref{Fig:HammingvsIpsenZOOM}(c) is detailed for subnetwork inference method $\mathcal{N}$.

After retaining the most distant pathways, we performed a comparison of common terms for fixed selection method $\mathcal{M}$ and database $\mathcal{D}$. The results are reported in Table \ref{Tab:NPathway_0p05}.
\begin{table}[!t]
\processtable{Summary of common most disrupted pathways ($\phi \ge 0.05$).\label{Tab:NPathway_0p05}}
{\begin{tabular}{cccc}\toprule
$\mathcal{M}$ & $\mathcal{D}$ & \multicolumn{1}{c}{$\left |\cap(\text{All}_\mathcal{E})\right |$}& \multicolumn{1}{c}{$\left |\cap(\mathcal{E}_{\text{WG, PaLS}})\right |$}\\\midrule
\multirow{2}{*}{$\ell_1\ell_2$}& GO   & 0 & 17  \\
                               			   & KEGG & 1  & 22 \\
\multirow{2}{*}{Liblinear}      & GO   & 0 & 5\\
                                            & KEGG & 0  & 21 \\\botrule
\end{tabular}}{}
\end{table}
In Tables \ref{Tab:GOCommonPathways_0p05} and \ref{Tab:KEGGCommonPathways_0p05} we report the most disrupted GO terms and KEGG pathways that have a {\em glocal} distance $\phi$ greater or equal to the chosen threshold $\tau$.

\begin{table*}[!t]
\processtable{Summary of  most disrupted GO terms common between WG and PaLS,  for different models $\mathcal{M}$ . Each  GO term  is associated to a {\em glocal} distance $\phi \ge 0.05$ for all subnetwork reconstruction algorithms $\mathcal{N}$. 
GO terms are sorted according decreasing average $\phi$. Bold fonts represent the GO terms shared by model $\mathcal{M}$.\label{Tab:GOCommonPathways_0p05}}
{\begin{tabular*}{\textwidth}{cp{.37\textwidth}|cp{.4\textwidth}}\toprule
\multicolumn{2}{c}{$\ell_1\ell_2$} & \multicolumn{2}{c}{Liblinear}\\\midrule
ID&\multicolumn{1}{c}{Term name}&\multicolumn{1}{c}{ID}&\multicolumn{1}{c}{Term name}\\\midrule
GO:0005739&Mitochondrion                                               & GO:0042127&Regulation of cell proliferation\\
GO:0031966&Mitochondrial membrane                                      & GO:0005783&Endoplasmic reticulum\\
GO:0005743&Mitochondrial inner membrane                                & GO:0015629&Actin cytoskeleton\\
GO:0042802&Identical protein binding                                   & GO:0006469&Negative regulation of protein kinase activity\\
GO:0007018&Microtubule-based movement                                  & {\bf GO:0005747}&{\bf Mitochondrial respiratory chain complex I}\\
GO:0046961&Proton-transporting ATPase activity, rotational mechanism   &&\\
GO:0005753&Mitochondrial proton-transporting ATP synthase complex      &&\\
GO:0000502&Proteasome complex                                          &&\\
GO:0015986&ATP synthesis coupled proton transport                      &&\\
GO:0045202&Synapse                                                     &&\\
GO:0048487&Beta-tubulin binding                                        &&\\
GO:0042734&Presynaptic membrane                                        &&\\
{\bf GO:0005747}&{\bf Mitochondrial respiratory chain complex I}       &&\\
GO:0006120&Mitochondrial electron transport, NADH to ubiquinone        &&\\
GO:0015078&Hydrogen ion transmembrane transporter activity             &&\\
GO:0015992&Proton transport                                            &&\\
GO:0005874&Microtubule                                                 &&\\\botrule
\multicolumn{4}{p{0.9\textwidth}}{}\\
\end{tabular*}}{}
\end{table*}

\begin{figure*}[tb]%
\begin{center}
\begin{tabular}{ccc}
\includegraphics[width=0.14\textwidth]{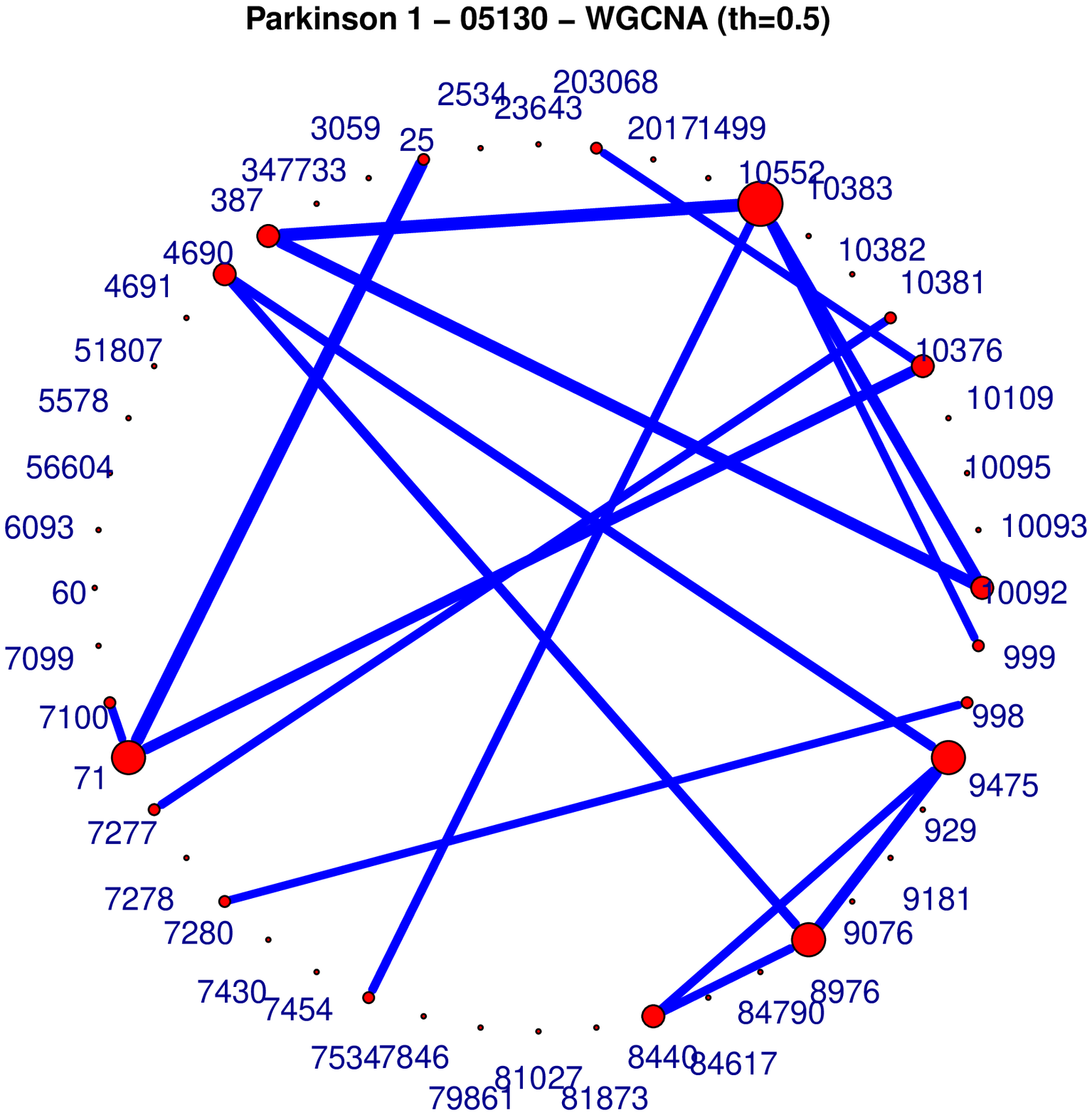} &
\multirow{2}{*}[65pt]{{\includegraphics[angle=-90,width=0.32\textwidth]{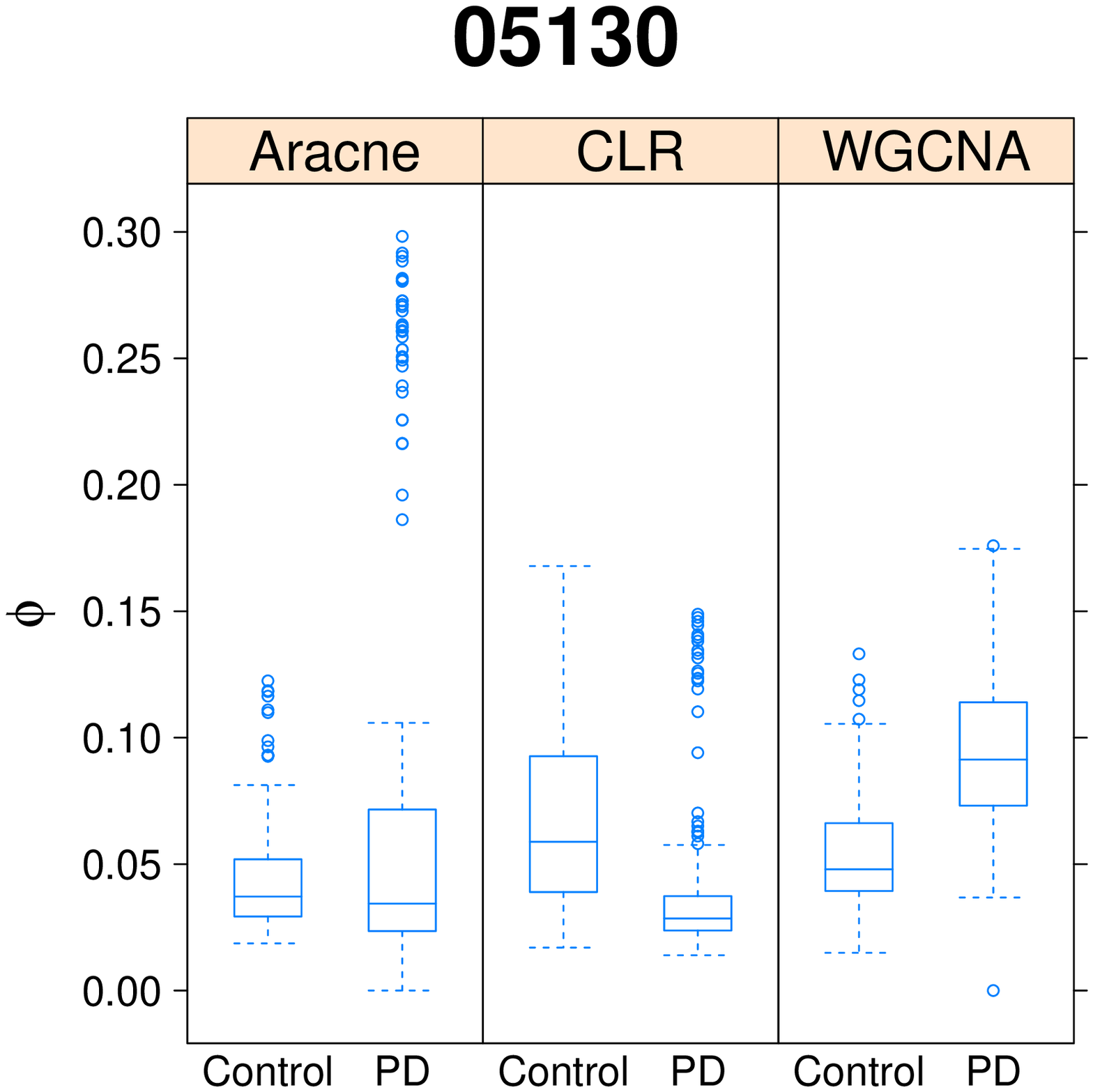}}}&
\includegraphics[width=0.14\textwidth]{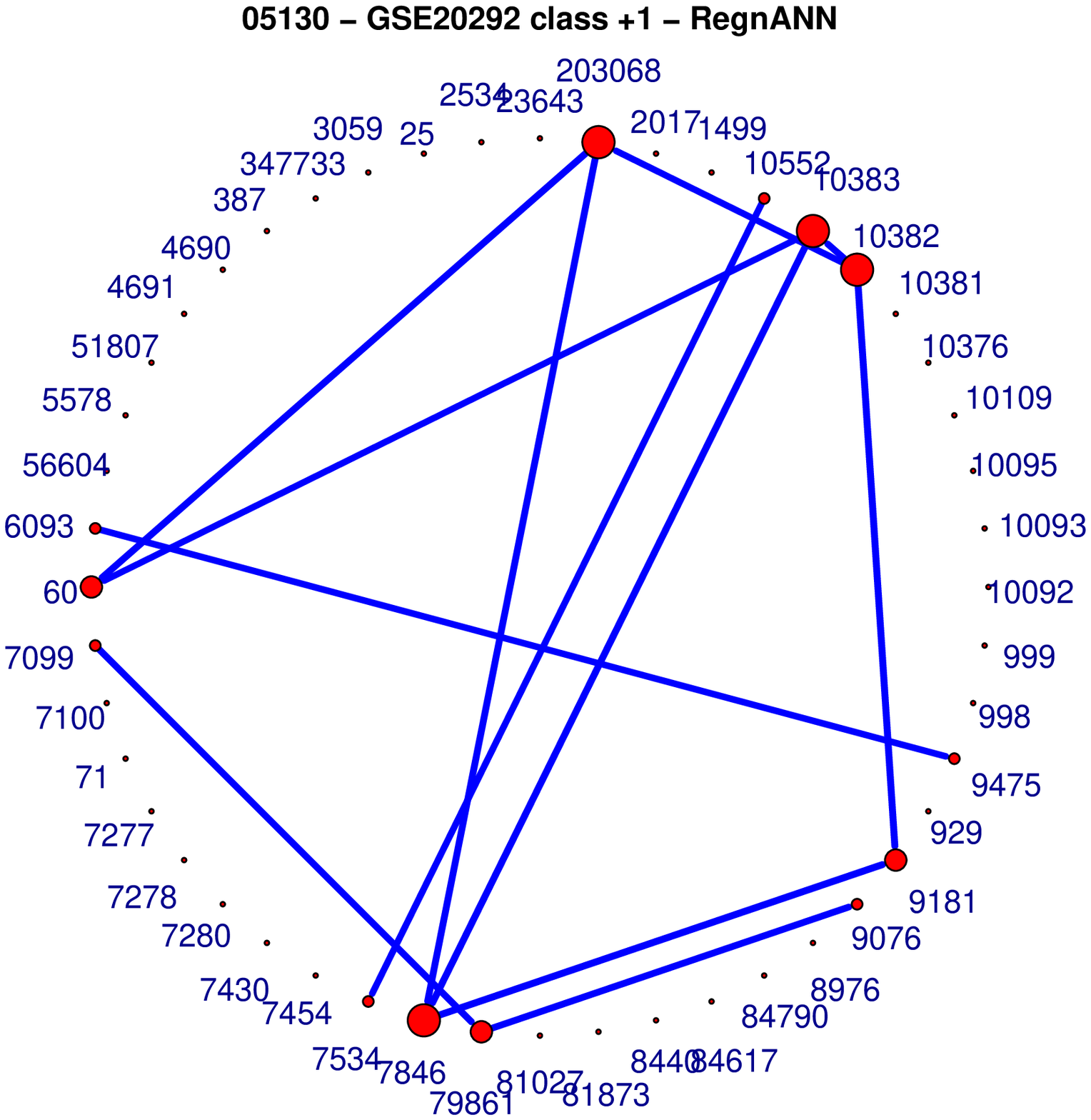}\\
\includegraphics[width=0.14\textwidth]{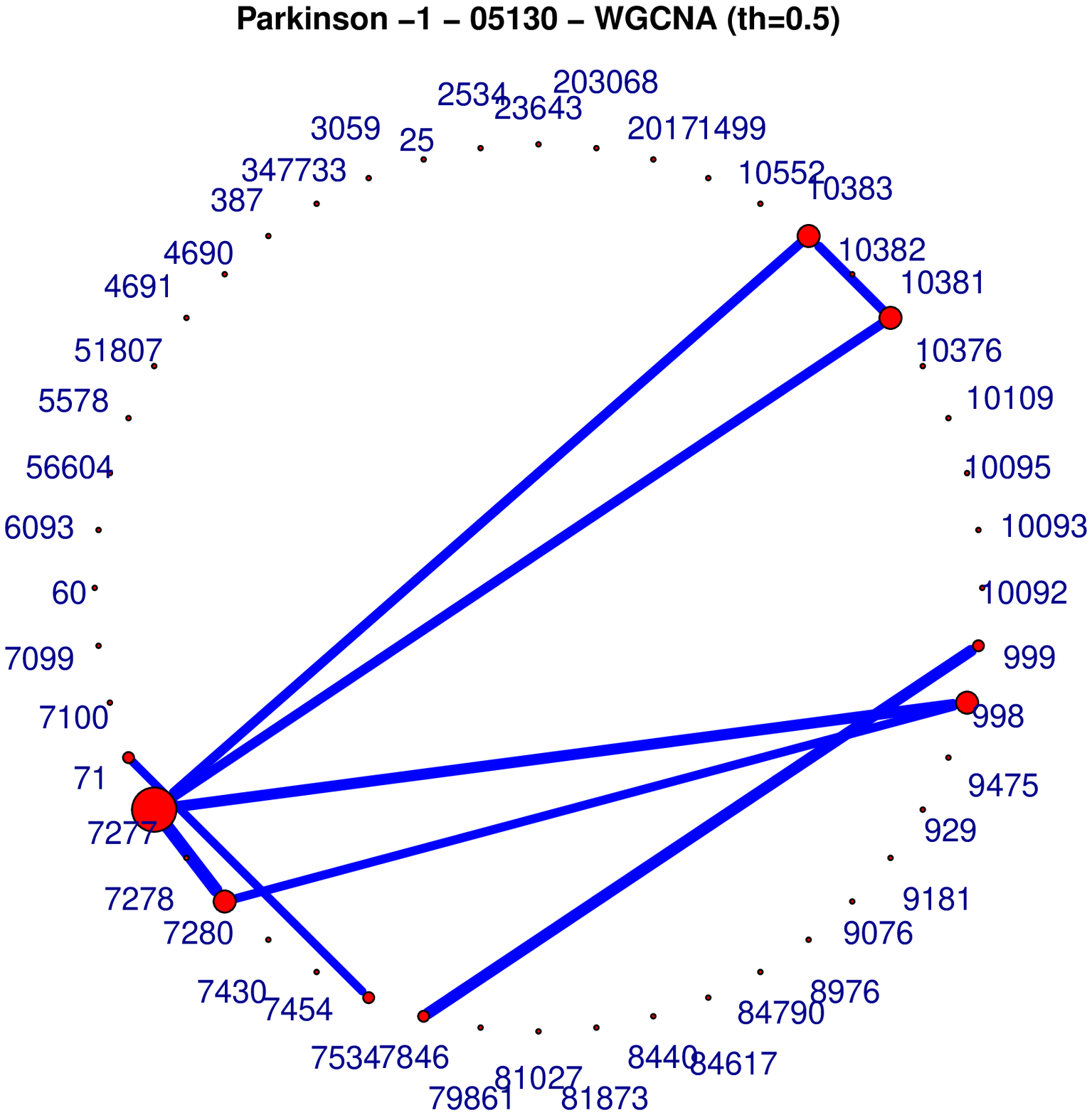} &
&
\includegraphics[width=0.14\textwidth]{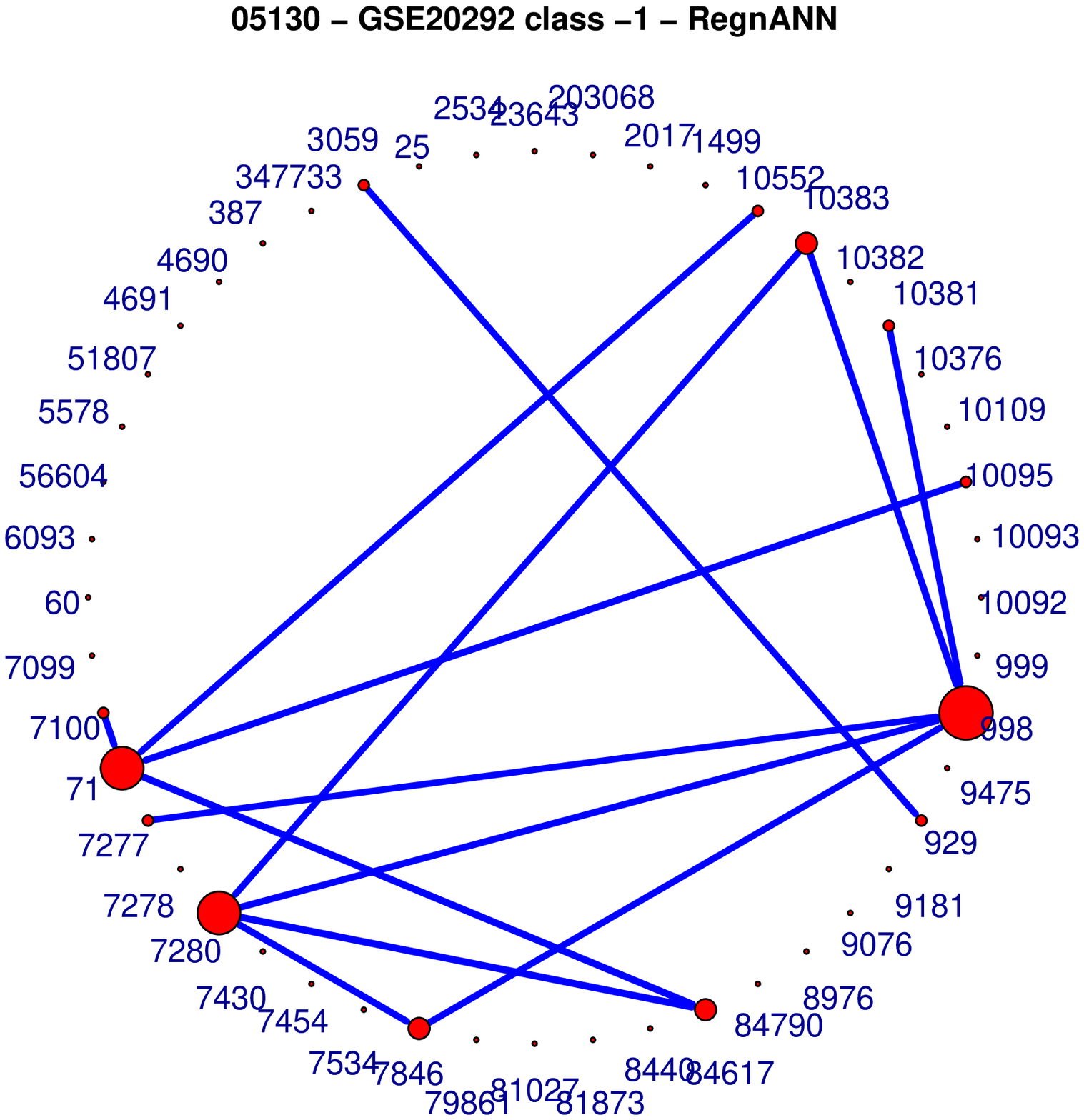}\\
(a) & (b) & (c)\\
\end{tabular}
\end{center}
\caption{(a) Networks inferred by WGCNA algorithm for the {\em Pathogenic E. coli infection} KEGG pathway for
         PD patients (above) and controls (below), 
         on the same pathway for different inference algorithm. (b) WGCNA is the
         method showing the highest stability on the two classes. (c) Same pathway reconstructed with RegnANN.}
\label{Fig:Ecoli}
\end{figure*}

\begin{table*}[tb]
\processtable{Summary of  most disrupted KEGG pathways common between WG and PaLS,  for different models $\mathcal{M}$. Each pathway is associated to a {\em glocal} distance $\phi \ge 0.05$  for all subnetwork reconstruction algorithms $\mathcal{N}$.
KEGG pathways are sorted according decreasing average $\phi$. Bold fonts represent the KEGG pathways shared by model $\mathcal{M}$.
\label{Tab:KEGGCommonPathways_0p05}}
{\begin{tabular*}{\textwidth}{cp{.416\textwidth}|cp{.4\textwidth}}\toprule
\multicolumn{2}{c}{$\ell_1\ell_2$} & \multicolumn{2}{c}{Liblinear}\\\midrule
ID& \multicolumn{1}{c}{Pathway name} &ID& \multicolumn{1}{c}{Pathway name}\\\midrule
{\bf 01100}&{\bf Metabolic pathway}                                    &  04630&Jak-STAT signaling pathway\\
{\bf 05130}&{\bf Pathogenic Escherichia coli infection}                &  {\bf 01100}&{\bf Metabolic pathway}\\
{\bf 04910}&{\bf Insulin signaling pathway}                            &  {\bf  05130}&{\bf Pathogenic Escherichia coli infection}\\
00310&Lysine degradation                                               &  04623&Cytosolic DNA-sensing pathway\\
04140&Regulation of autophagy                                          &  00330&Arginine and proline metabolism\\
03050&Proteasome                                                       &  {\bf 04910}&{\bf Insulin signaling pathway}\\
00230&Purine metabolism                                                &  05212&Pancreatic cancer\\
05014&Amyotrophic lateral sclerosis*                                   &  03030&DNA replication\\
00980&Metabolism of xenobiotics by cytochrome P450                     &  {\bf 05213}&{\bf Endometrial cancer}\\
00620&Pyruvate metabolism                                              &  04660&T cell receptor signaling pathway\\
{\bf 05213}&{\bf Endometrial cancer}                                   &  04310&Wnt signaling pathway\\
00270&Cysteine and methionine metabolism                               &  05210&Colorectal cancer\\
00240&Pyrimidine metabolism                                            &  04912&GnRH signaling pathway\\
05120&Epithelial cell signaling in Helicobacter pylori infection       &  05332&Graft-versus-host disease\\
05110&Vibrio cholerae infection                                        &  04520&Adherens junction\\
00020&Citrate cycle (TCA cycle)                                        &  04621&NOD-like receptor signaling pathway\\
00562&Inositol phosphate metabolism                                    &  04370&VEGF signaling pathway\\
00600&Sphingolipid metabolism                                          &  04662&B cell receptor signaling pathway\\
05218&Melanoma                                                         &  {\bf  04722}&{\bf Neurotrophin signaling pathway}\\
00010&Glycolysis / Gluconeogenesis                                     &  05214&Glioma\\
00051&Fructose and mannose metabolism                                  &  04330&Notch signaling pathway\\
{\bf 04722}&{\bf Neurotrophin signaling pathway}                       & &\\\midrule
\multicolumn{4}{p{0.9\textwidth}}{\textit{\small{*Note: This is the only selected pathway shared across all enrichment methods $\mathcal{E}$. }}}\\\botrule
\end{tabular*}}{}
\end{table*}
\begin{figure*}[tb]%
\begin{center}
\begin{tabular}{ccc}
\includegraphics[width=0.14\textwidth]{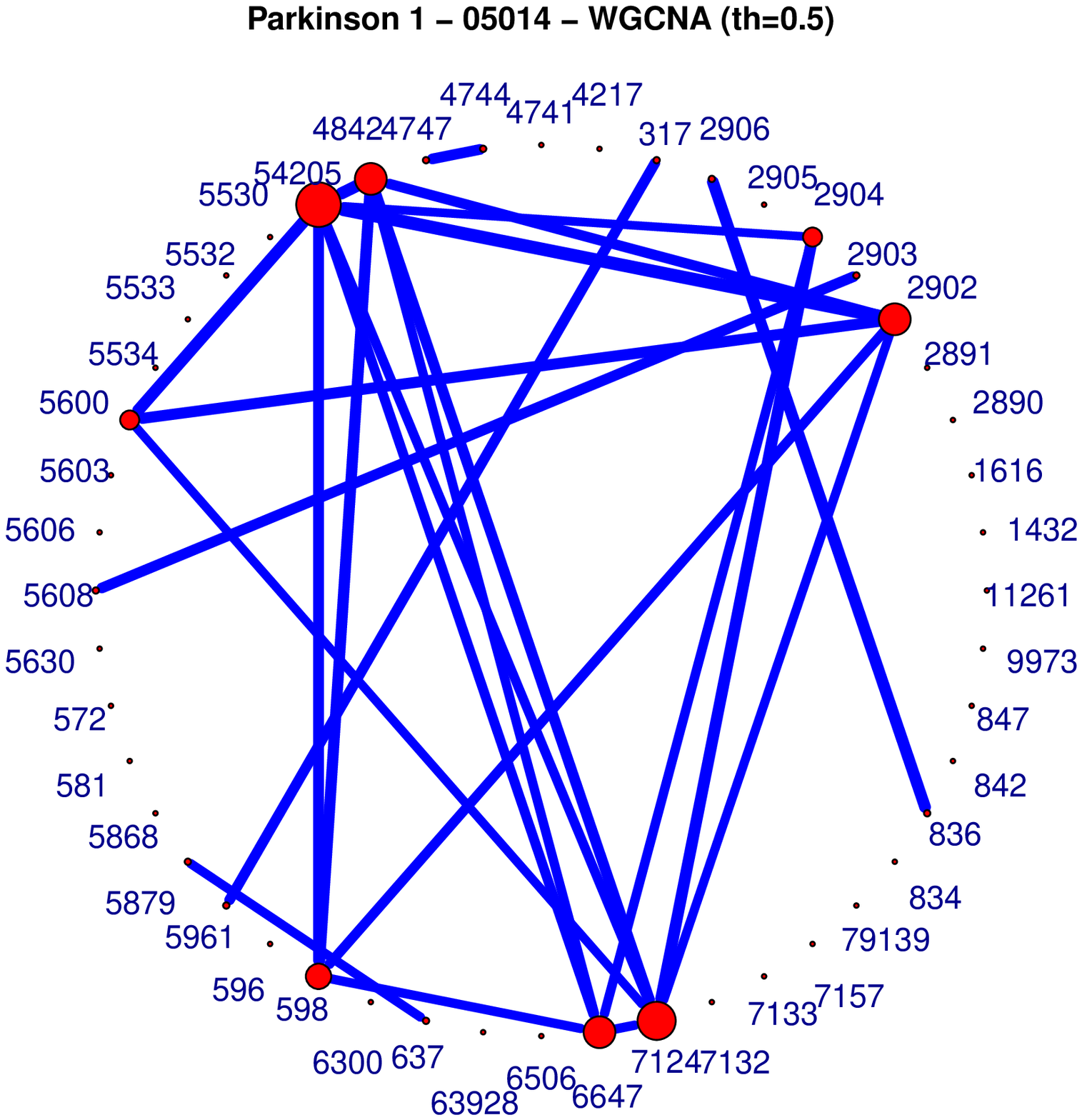} &
\multirow{2}{*}[65pt]{{\includegraphics[angle=-90,width=0.32\textwidth]{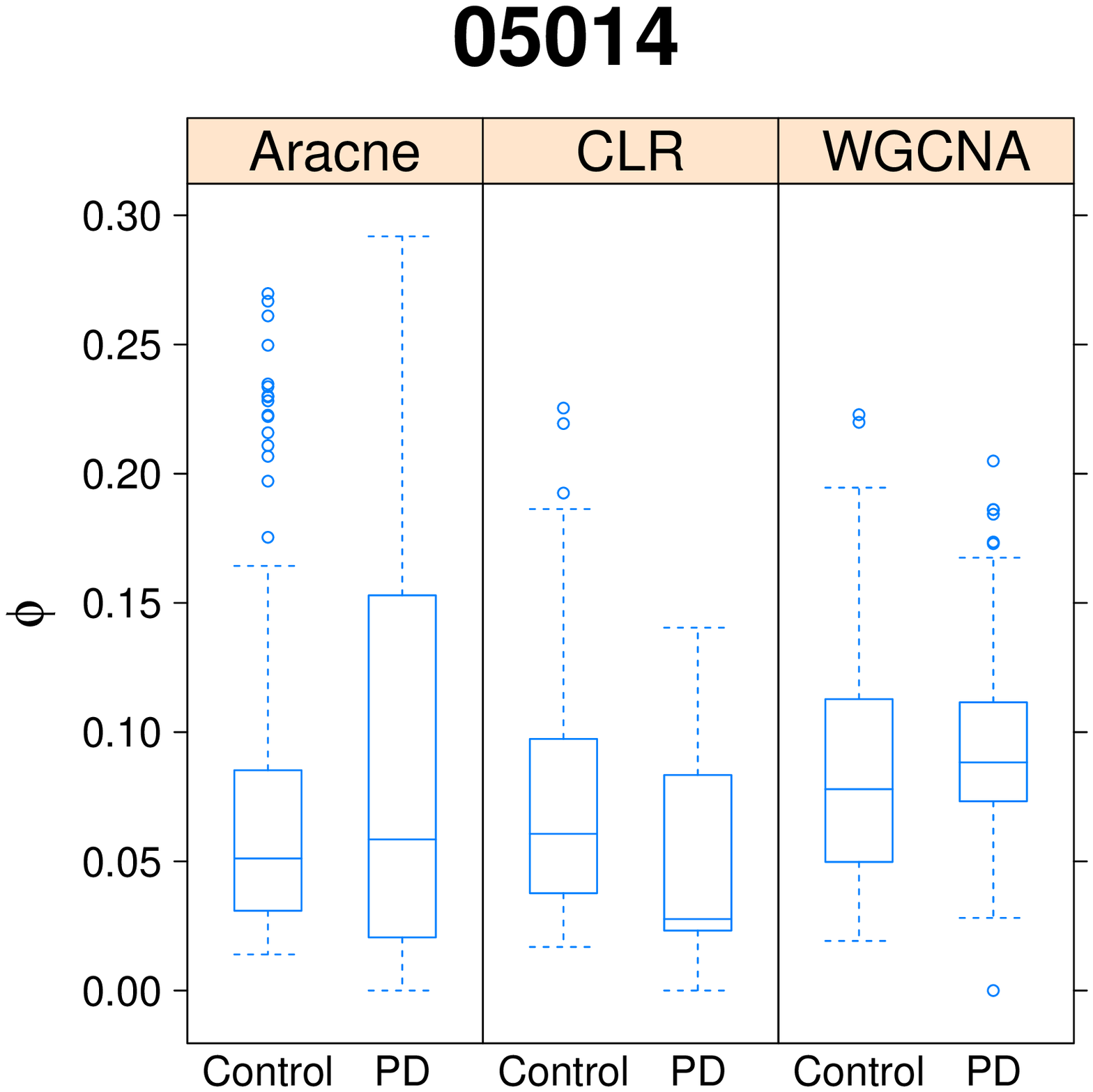}}}&
\includegraphics[width=0.14\textwidth]{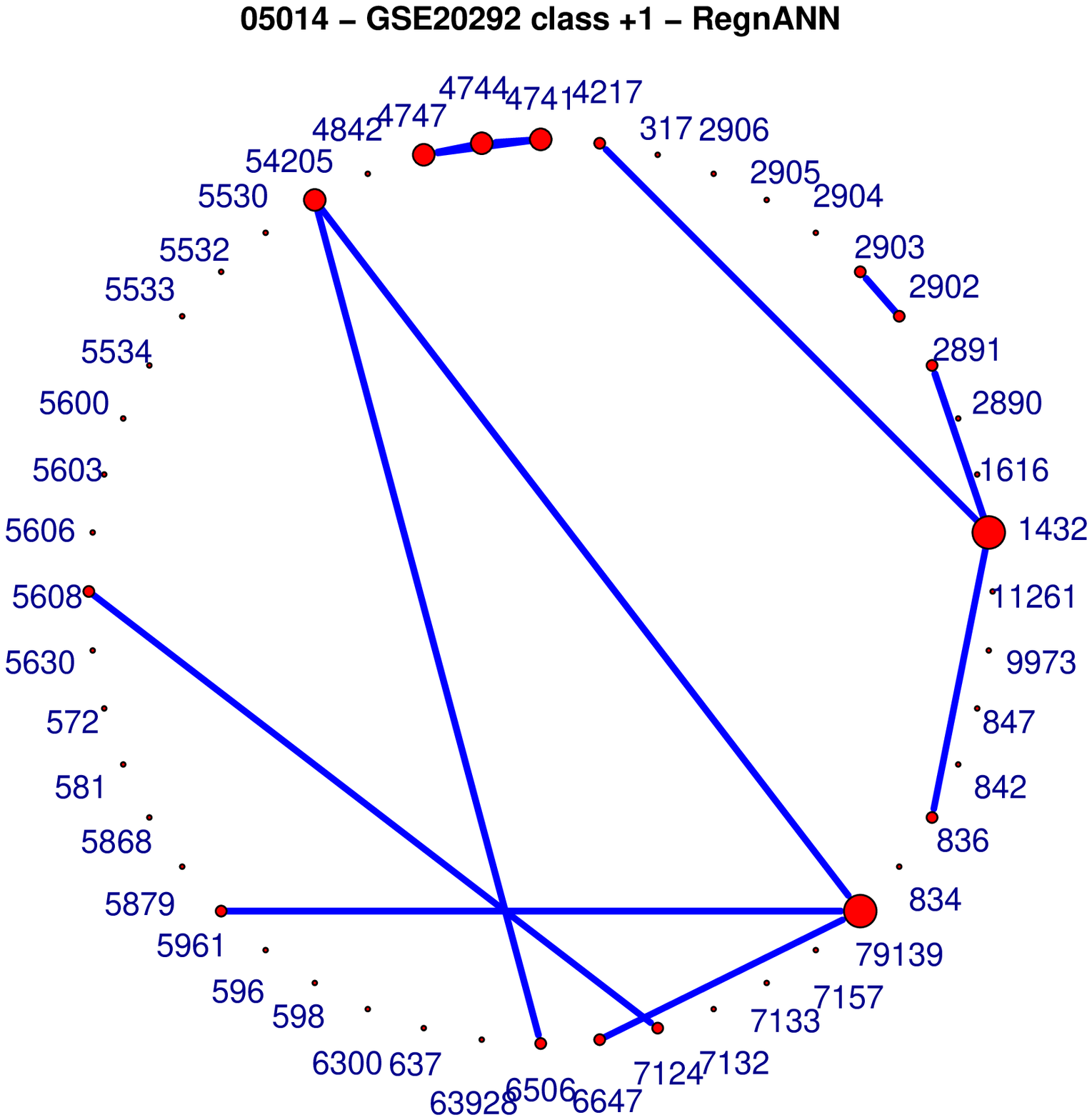}\\
\includegraphics[width=0.14\textwidth]{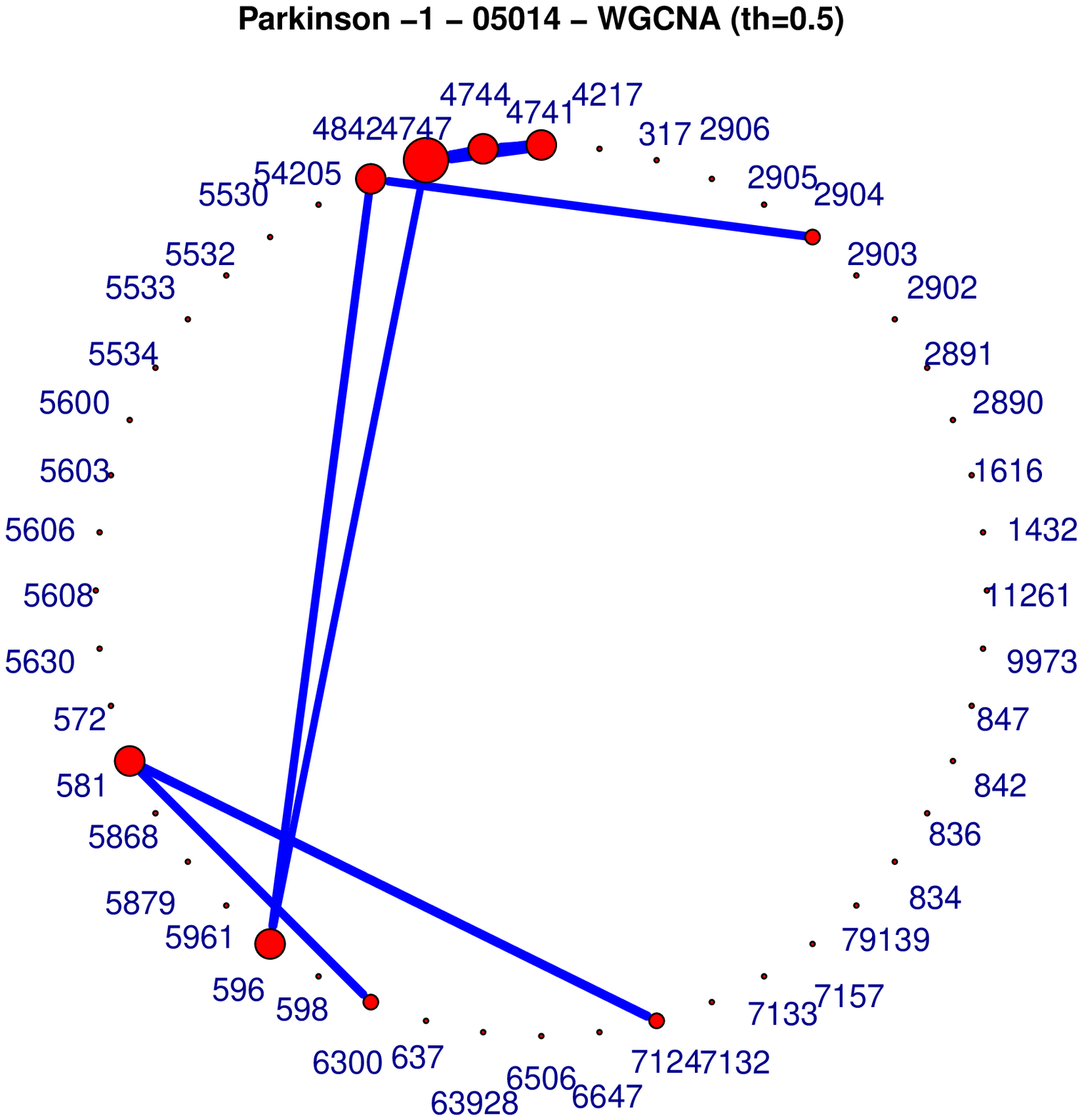} &
&
\includegraphics[width=0.14\textwidth]{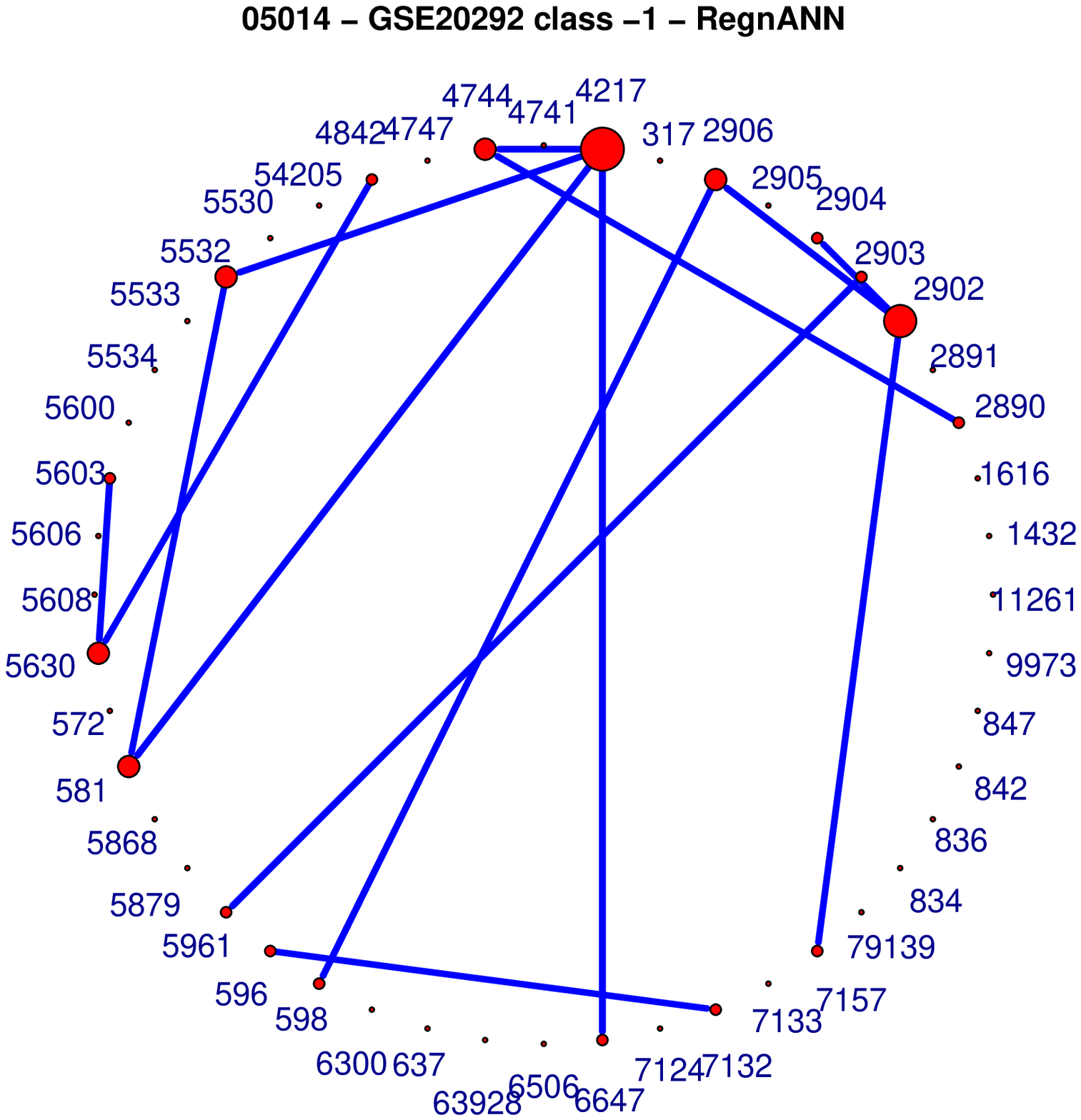}\\
(a) & (b) & (c)\\
\end{tabular}
\end{center}
\caption{(a) Networks inferred by WGCNA algorithm for the {\em ALS} KEGG pathway for
         PD patients (above) and controls (below), 
         on the same pathway for different inference algorithm. (b) WGCNA is the
         method showing the highest stability on the two classes. (c) Same pathway reconstructed with RegnANN.}
\label{Fig:ALS}
\end{figure*}
As an example of a selected pathway within KEGG, 
the networks (thresholded at edge weight 0.1 for graphic purposes) inferred by WGCNA (together with the
corresponding stability) on the {\em Amyotrophic Lateral Sclerosis} KEGG pathway (ALS - 05014) are displayed in
Figure~\ref{Fig:ALS}. 
We also plot the inferred network by  the  RegnANN algorithm.
Similarly, in Figure \ref{Fig:Ecoli} we plot the  {\em Pathogenic E. coli infection} KEGG pathway, reconstructed by WGCNA, its stability plot, and the corresponding inferred networks by  the  RegnANN algorithm. 

\begin{figure*}[th]%
\begin{center}
\begin{tabular}{ccc}
\includegraphics[width=0.4\textwidth,angle=270]{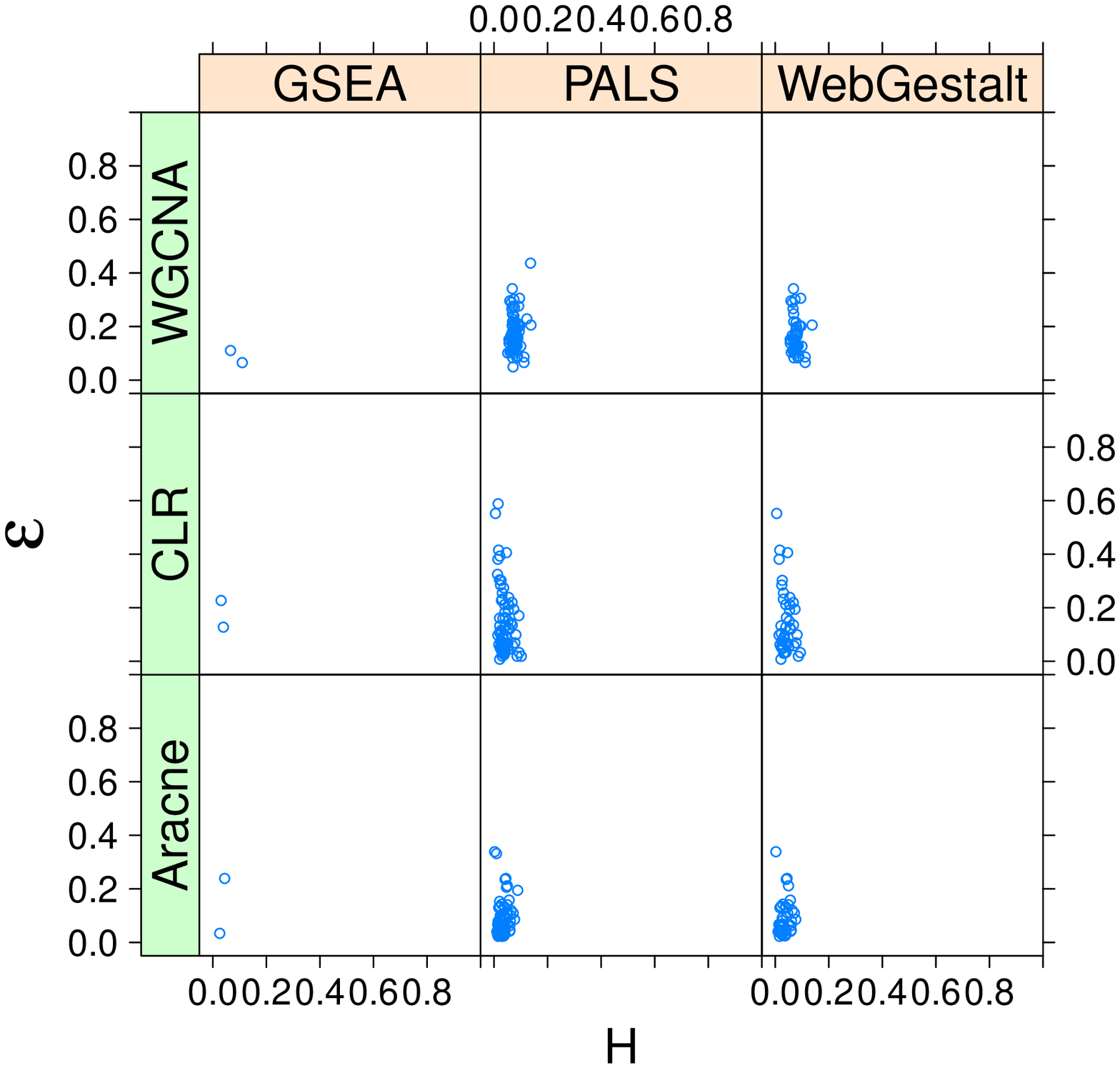} &
&
\includegraphics[width=0.4\textwidth,angle=270]{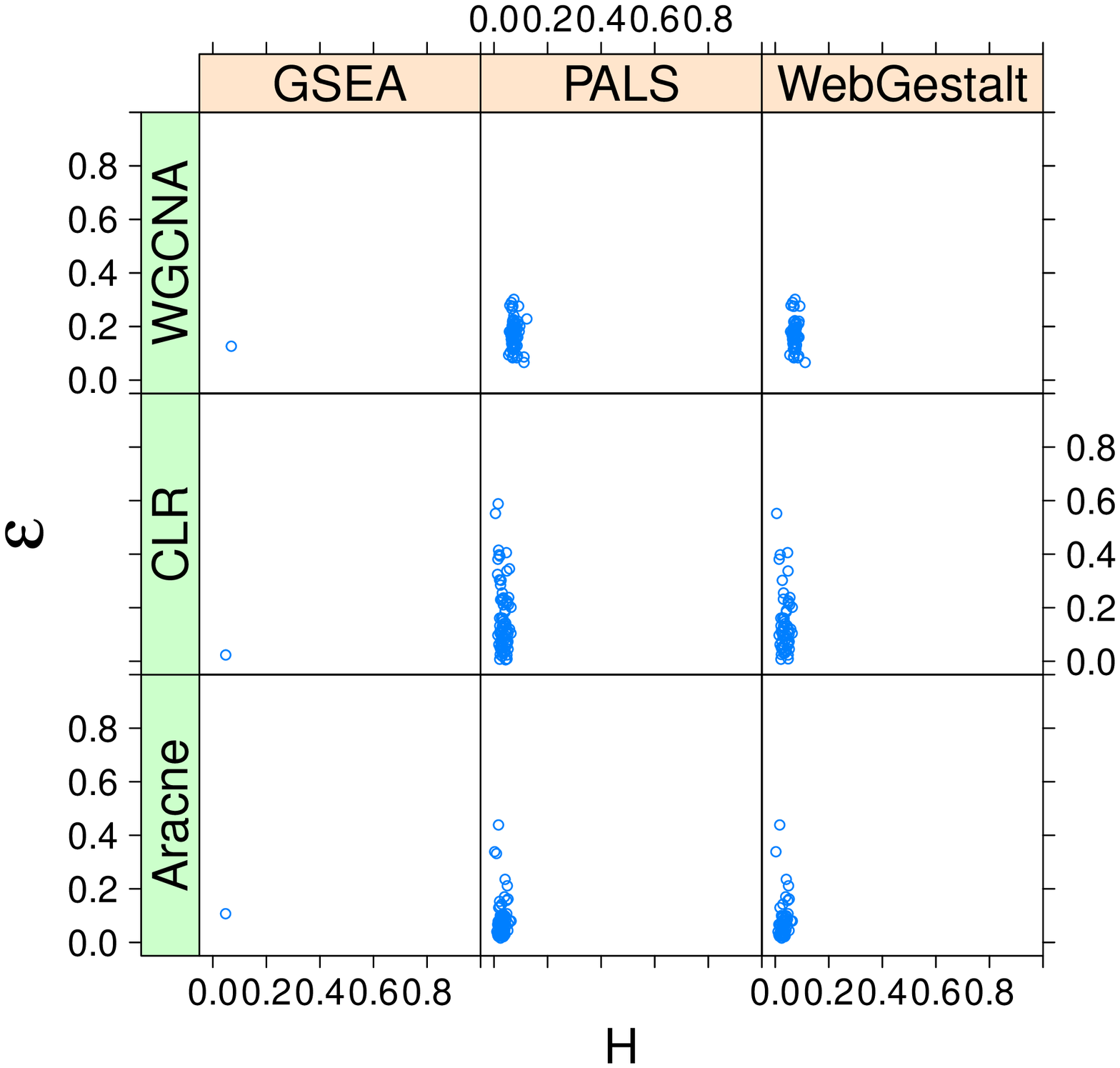} \\
a: $\ell_1\ell_2$ and KEGG && b: Liblinear and KEGG\\
\includegraphics[width=0.4\textwidth,angle=270]{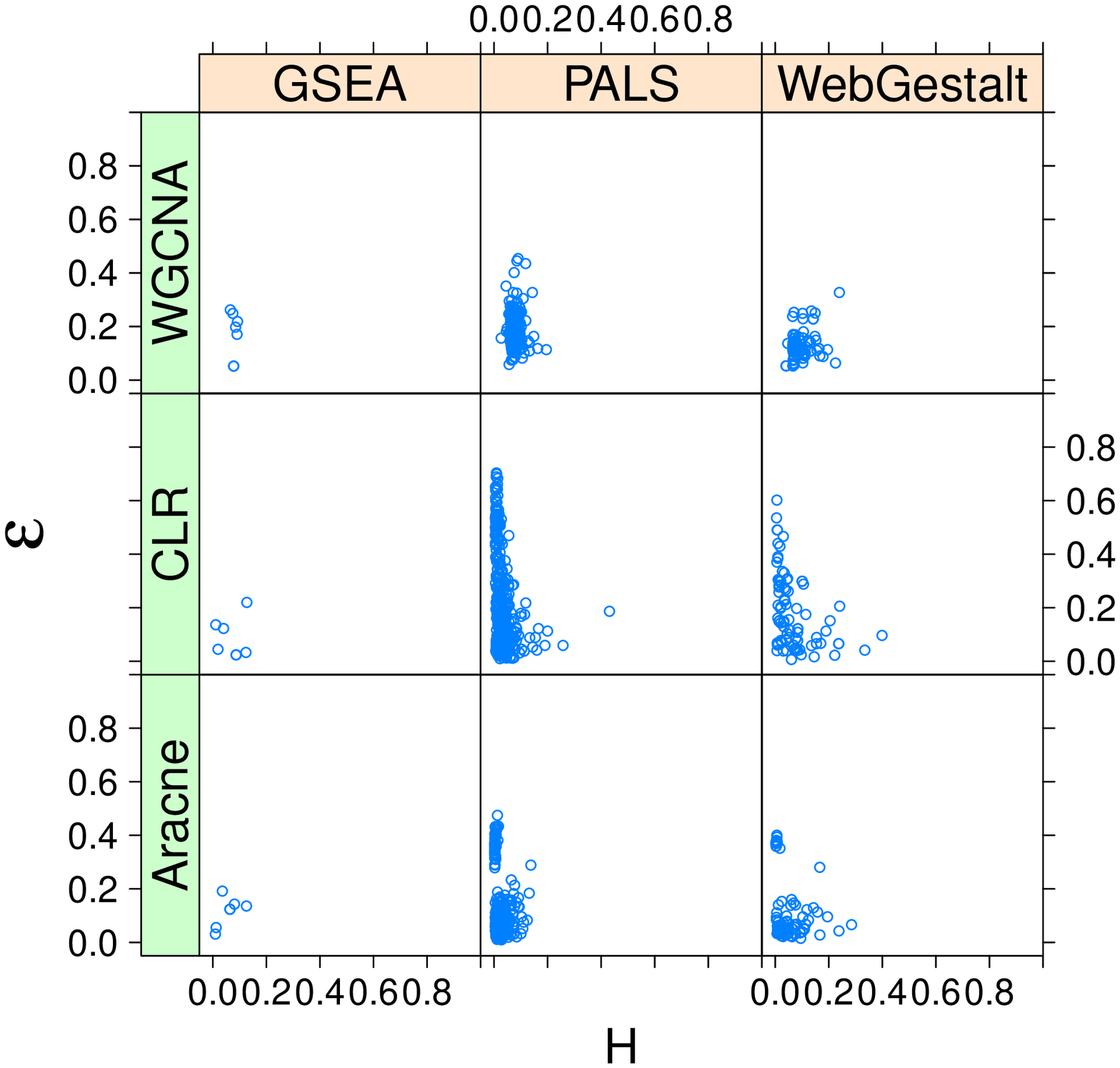} &
&
\includegraphics[width=0.4\textwidth,angle=270]{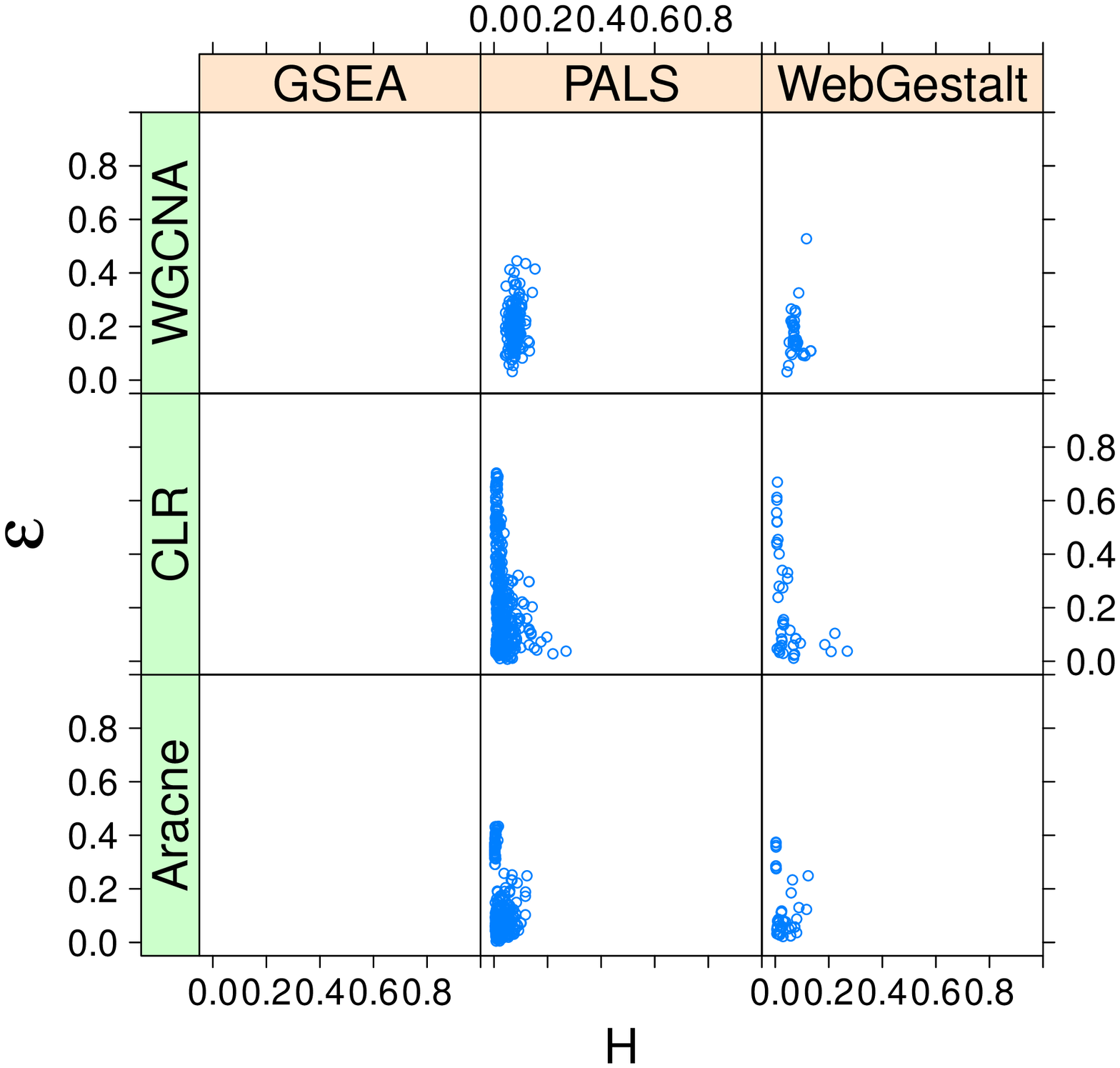}\\
c: $\ell_1\ell_2$ and GO && d: Liblinear and GO
\end{tabular}
\end{center}
\caption{Plots of Hamming vs. Ipsen distances (H vs. $\epsilon$) for all possible combinations of $\mathcal{M}, \mathcal{D},  \mathcal{E}$  and $\mathcal{N}$.
In our analysis we considered the {\em glocal} distance $\phi$, defined as the normalized product of H and $\epsilon$.
}
\label{Fig:HammingvsIpsen}
\end{figure*}

\noindent {\bf Discussion}\noindent \\
The variability in the results, as expected, strongly depends on the method of choice. 
For feature selection, the nature of the method is key. In the proposed pipeline we limited the impact of this step by choosing two approaches within the regularization methods family. Both classifiers adopt a $\ell_2$-regularization penalty term, combined with different loss functions and, for $\ell_1\ell_2$ with another regularization term. 
We used similar but not equal model selection protocols. Both guarantee that the results are not affected by selection-bias.
In this work, the main source of variability  was the choice of the gene enrichment module.  Therefore, the experimenter 
must be careful in choosing one method or another and in using it compliantly with the experimental design. 
For instance, GSEA was designed for 
estimating the significance levels by considering separately the positively and negatively scoring 
gene sets within a list of genes selected with {\em filter} methods based on classical statistical tests.
It is worth noting that, if one uses the preranked option, as we did, negative regulated groups might not be significant at all (we indeed discarded them). 
WG uses the Hypergeometrical test to assess the functional groups but, differently from GSEA, does not use any significance assessment based on permutation of phenotype labels.
PaLS is the simplest methods, being just a measure of occurrences of a given descriptor in the list of selected genes.
However, enrichment methods from different categories are complementary and can identify different but equally
meaningful biological aspects of the same process.
Thus, the integration of information across different methods is the best strategy. 

Moreover, the assessment of the reconstruction distance between case and control version of the same pathways help in providing a quantitative focus on the key pathway involved in the process. The use of a distance mixing the effects of structural changes with those due to the differences in rewiring moreover warrants a more informative view on the difference assessment itself.
The limited effect of different feature selection methods is confirmed by the plots in Figure~\ref{Fig:HammingvsIpsen}.

For  $\ell_1\ell_2$, the only most disrupted pathway  shared by the three enrichment tools $\mathcal{E}$ and 
the three reconstruction methods  $\mathcal{N}$ is ALS.
This pathway is relevant in this context because, like PD, ALS  is another neurodegenerative disease
therefore they share significant biological features in particular at the mithocondrial level. Moreover
at the phenotypic level the skeletal muscles of the patients are severely affects influencing
the movements. 
In Figure~\ref{Fig:ALS}  it is evident that a high number of interactions
are established among the genes going from the control (below) to the affected (above) pathways.
It is also interesting to underline that CYCS (Entrez ID: 54205) one of the hub genes (represented by a red dot in the graph) 
within the pathway  was identified by $\ell_1\ell_2$ as discriminant. 
This gene is highly involved in several neurodegenerative diseases ({\em e.g.,}
PD, Alzheimer's, Huntington's) and in pathways related to cancer. Furthermore its protein is known to functions
as a central component of the electron transport chain in mitochondria and to be involved in initiation of apoptosis,
known cause of the neurons loss in PD.
Across variable selection algorithms  $\mathcal{M}$, five highly disrupted pathways were found as shared between two of the three enrichment methods (see Table \ref{Tab:KEGGCommonPathways_0p05}, bold items). 
In particular, we represented in Figure~\ref{Fig:Ecoli} the corresponding inferred networks. 
To further highlight the different outcomes occurring from the same dataset when diverse inference methods are employed, we reconstructed the {\em ALS} and {\em Pathogenic E. coli infection} by the RegnANN algorithm, which tends to spot also second order correlation among the network nodes, see Figures~\ref{Fig:ALS} and \ref{Fig:Ecoli}.

Two genes in the {\em E. coli infection} pathway were selected both by $\ell_1\ell_2$ and Liblinear, namely ABL1 (Entrez ID:71) and TUBB6 (Entrez ID: 84617).  ABL1 seems to play a relevant role as hub both in the WGCNA and in the RegnANN networks. ABL1 is a protooncogene that encodes protein tyrosine kinase that has been implicated in processes of cell differentiation, cell division, cell adhesion, and stress response. It was also found to be responsible of apoposis in human brain microvascular endothelial cells.
\begin{figure}[tb]%
\begin{center}
\begin{tabular}{c}
\includegraphics[width=0.4\textwidth]{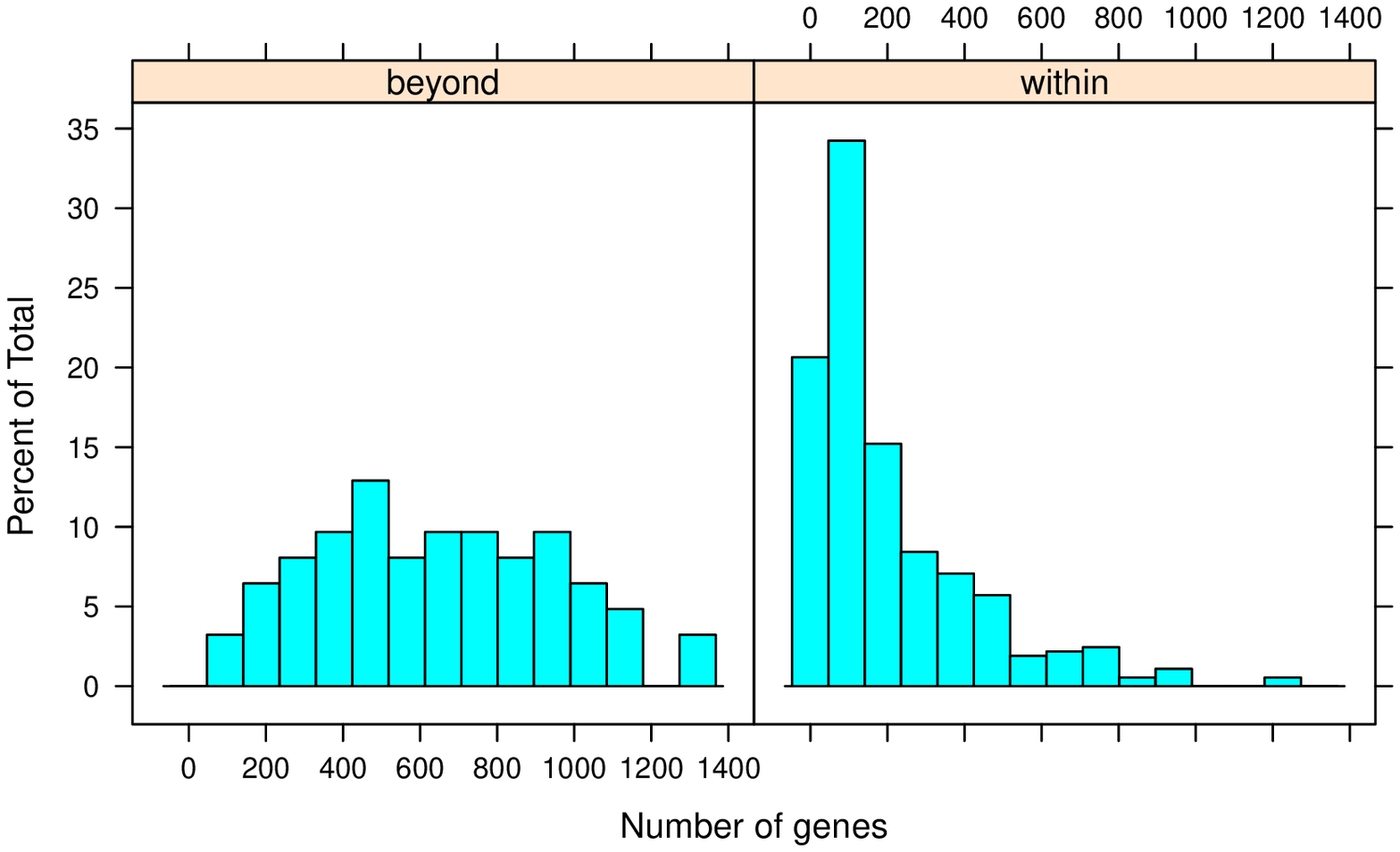}\\
(a)\\
\includegraphics[width=0.4\textwidth]{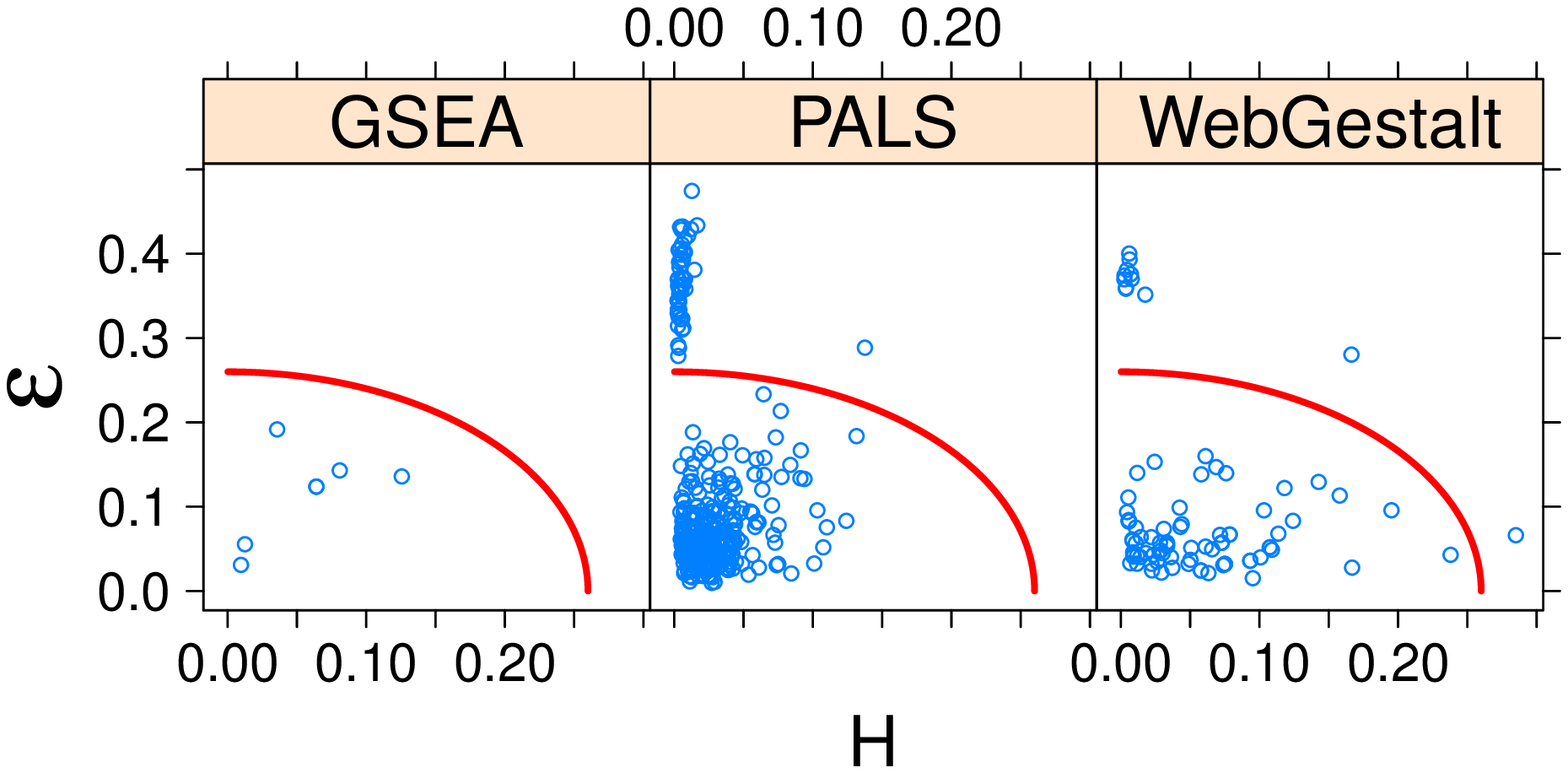}\\
(b)\\
\end{tabular}
\end{center}
\caption{(a) Pathway target cumulative histogram. (b) Hamming versus Ipsen (H vs. $\epsilon$) distance, and thresholding of high populated pathways.}
\label{Fig:HammingvsIpsenTHR}
\end{figure}
In Figure~\ref{Fig:HammingvsIpsenTHR} we note that pathways with high number of genes are similar in term of local distance, instead a wider range of variability is found looking at the spectral distance. The red line in \ref{Fig:HammingvsIpsenTHR}(b) divides the 2 cluster. Pathway targets beyond and within the red line are represented in the cumulative histogram in  \ref{Fig:HammingvsIpsenTHR}(a). Pathways beyond the threshold are equally distributed and they represent a wider range of targets, instead pathways within the threshold show a smaller number of targets  \ref{Fig:HammingvsIpsenTHR}(a) on the right.

\section{Conclusion}
Moving from gene profiling towards pathway profiling can be an effective solution to overcome the problem of the poor overlapping in {\em -omics} signatures.
Nonetheless, the path from translating a discriminant gene panel into a coherent set of functionally related gene sets includes a number of steps each contributing in injecting variability in the process.
To reduce the overall impact of such variability, it is thus critical that, whenever possible, the correct tool for each single step is adopted, accurately focussing on the desired target to be investigated.
This mainly holds for the choice of the most suitable enrichment tool and biological knowledge database, and, to a lower extent, to the inference method for the newtork reconstruction: all these ingredients are planned for different objectives, and their use on other situations may result misleading.
As a final observation and a possible future development to explore, the emerging instability can be tackled by obtaining the functional groups identification as the result of a prior knowledge injection in the learning phase, rather than a procedure a posteriori \citep{Zycinski:2011, Zycinski:2012}.

\section*{Acknowledgement}
The authors at FBK  want to thank Shamar Droghetti for his help with the
enrichment web interfaces.

\paragraph{Funding\textcolon}
The authors at DISI acknowledge funding by the Compagnia di San Paolo funded Project {\em Modelli e metodi computazionali nello studio della fisiologia e patologia di reti molecolari di controllo in ambito oncologico}.
The authors at FBK acknowledge funding by the European Union FP7 Project HiperDART and by the PAT funded Project ENVIROCHANGE.

\bibliographystyle{natbib}
\bibliography{barla12evaluating}
\end{document}